\documentclass[12pt]{article}

\usepackage{amssymb}
\usepackage{amsmath}
\usepackage{amsfonts}
\usepackage{graphicx}
\usepackage{subfigure}
\usepackage{color}
\usepackage{dcolumn}
\usepackage{hyperref}
\numberwithin{equation}{section}
\newcommand{\be}{\begin{equation}}
\newcommand{\ee}{\end{equation}}
\newcommand{\bea}{\begin{eqnarray}}
\newcommand{\eea}{\end{eqnarray}}
\newcommand{\nn}{\nonumber}
\newcommand{\rd}{\partial}

\begin{document}
\begin{titlepage}
\thispagestyle{empty}

\hfill  WITS-CTP-99

\vspace{2cm}

\begin{center}
\font\titlerm=cmr10 scaled\magstep4 \font\titlei=cmmi10
scaled\magstep4 \font\titleis=cmmi7 scaled\magstep4 {
\Large{\textbf{Energy loss in a strongly coupled anisotropic plasma}
\\}}
\vspace{1.5cm}
\noindent{{%\large
Kazem Bitaghsir Fadafan$^{a}$\footnote{e-mail:bitaghsir@shahroodut.ac.ir },
Hesam Soltanpanahi$^{b}$\footnote{e-mail:hesam.soltanpanahisarabi@wits.ac.za }
}}\\
\vspace{0.8cm}

{\it
${^a}$Physics Department, Shahrood University of Technology, Shahrood, Iran\\}

\vspace*{.4cm}

{\it ${}^b$National Institute for Theoretical Physics, Gauteng\\
School of Physics and Center for Theoretical Physics,\\
University of the Witwatersrand,\\ WITS 2050, Johannesburg, South Africa}

\end{center}

\vskip 2em

%----------------------------------------------------------------------------
\begin{abstract}
We study the energy loss of a rotating infinitely massive quark
moving, at constant velocity, through an anisotropic strongly-
coupled $\mathcal{N} = 4$ plasma from holography. It is shown that,
similar to the isotropic plasma, the energy loss of the rotating
quark is due to either the drag force or radiation with a continuous
crossover from drag-dominated regime to the radiation dominated
regime. We find that the anisotropy has a significant effect on the
energy loss of the heavy quark, specially in the crossover regime.
We argue that  the energy loss due to radiation in anisotropic media
is less than the isotropic case. Interestingly this is similar to
analogous calculations for the energy loss in weakly coupled
anisotropic plasma.

\end{abstract}

\end{titlepage}

\tableofcontents
%----------------------------------------------------------------------------

%---------------------------------------------------------------------------------------------------------------------------------------------------
 \section{Introduction}
The experiments at Relativistic Heavy Ion Collisions (RHIC) and at
the Large Hadron Collider (LHC) have produced a strongly-coupled
quark$-$gluon plasma (QGP)(see review
\cite{CasalderreySolana:2011us}).
At a qualitative level, the data indicate that the QGP
produced at the LHC is comparably strongly-coupled \cite{ALICE}.
The QGP created at the LHC is
expected to be closer to conformality than the case at
RHIC \cite{CasalderreySolana:2011us}. The plasma created right after
a heavy ion collision is anisotropic for a period of time,
$\tau<\tau_{\text{iso}}$ and becomes locally isotropic afterwards
\cite{Romatschke:2003ms,Arnold:2005vb,Rebhan:2008uj}. There are no
known quantitative methods to study strong coupling phenomena in QCD
which are not visible in perturbation theory (except by lattice
simulations). A new method for studying different aspects of the QGP is
the $AdS/CFT$ correspondence
\cite{CasalderreySolana:2011us,Maldacena:1997re,Gubser:1998bc,Witten:1998qj,
Witten:1998zw}.

The energy loss of partons is a useful probe of the QGP. One should
notice that this quantity is not a direct experimental observable but
it is related to the jet quenching parameter $\hat{q}$. This
parameter is a signature of the QGP and initially was
calculated from AdS/CFT correspondence in \cite{Liu:2006ug}.
Although the energy loss can be due to various different mechanisms,
two primary sources of the energy loss are collisions and
radiation. The former is due to the scattering of the energetic
partons with thermal quarks and gluons in the QGP, while the later is
related to the Bremsstrahlung radiation during the interactions with
the medium. Theoretical study of these phenomena needs
strong-coupling tools and non-perturbative methods such as lattice
QCD, which are inadequate to describe these time-dependent phenomena.
A new method is to use the holographic principle and investigate the
energy loss of heavy probes.

This method has yielded many important insights into the dynamics of
strongly-coupled gauge theories.  It has been used to investigate
hydrodynamical transport quantities in various interesting
strongly-coupled gauge theories where perturbation theory is not
applicable. Using this method,  the energy loss due to collisions of
heavy quarks was initially discussed  in
\cite{Herzog:2006gh,Gubser:2006bz}. They considered a heavy quark
traveling with constant linear velocity through the plasma, and find
the energy required to keep it in uniform motion. This energy is
then given by the world-sheet momentum and falls across the horizon.
On the other hand, the energy loss is proportional to the momentum
of the heavy quark and one concludes that the energy loss mechanism
can be considered as a drag force. Some extension of this
calculation to more realistic models of QGP have been done in
\cite{Gubser:2009sn}. It should be noticed that the physically
relevant setting is when there is no external force acting on the
energetic parton. Then the moving probe is decelerating and here is
the possibility of another important mechanism of energy loss,
namely radiation
\cite{Dominguez:2008vd,Chernicoff:2008sa,Hatta:2011gh}. Therefore,
it is worthwhile to find interference between these two different
mechanisms of the energy loss,
 the drag force and radiation.

By considering the quark rotating uniformly with constant velocity
$v$, one can consider these two mechanisms of the energy loss at the
same time.\footnote{ The interplay between the energy loss due to
the radiation and the drag force can also be observed in the early
times energy loss of a heavy quark \cite{Guijosa:2011hf}. This study
has been done in the isotropic plasma, while one may expect that the
results will be affected by considering the anisotropy in the real
models.} The significant advantage offered by the analysis of this
problem was explored in \cite{Fadafan:2008bq}. They determined the
energy it takes to move a test quark along a circle of radius $l$
with angular frequency $\omega$ through the strongly-coupled plasma
of $\mathcal{N} = 4$ supersymmetric Yang-Mills (SYM) theory. The
drag force and radiation for fixed $v=l\omega$ are dominant when
($\omega\rightarrow 0,\,l\rightarrow \infty$) and
($\omega\rightarrow \infty,\,l\rightarrow 0$), respectively. The
extension of this calculation to the case of non-conformal plasma,
which exhibits a confinement/deconfinement transition at a critical
temperature, was done in \cite{AliAkbari:2011ue}. They argued that
in the low temperature limit the energy loss is completely due to
glueball Bremsstrahlung. The motion of a heavy charged quark in a
magnetic field was also analyzed in the vacuum of strongly-coupled
conformal field theory in \cite{Kiritsis:2011ha}.

In this paper, we study the energy loss of a rotating quark in a
strongly-coupled anisotropic plasma using the AdS/CFT
correspondence. Anisotropy is very important during the initial
phase after the QGP production in the heavy-ion collisions. It is
related to this point that initially the system expands mainly along
the collision axis. As a result, the plasma has significantly
unequal pressures in the longitudinal and transverse directions. To
model this medium, we consider an anisotropic $\mathcal{N} = 4$ SYM
plasma in which the $x, y$ directions are rotationally symmetric,
while the $z$ direction corresponds to the beam direction
\cite{Mateos:2011ix}. The drag force and jet quenching parameter in
this medium were calculated in
\cite{Chernicoff:2012iq,Giataganas:2012zy,Chernicoff:2012gu}. It was
found that the energy loss of partons depend on the relative
orientation between the anisotropic direction and the velocity of
the quark.

Another interesting feature of this anisotropic background is that
the ratio of the shear viscosity to the entropy density violated
KSS bound \cite{Kovtun:2004de}. While the ratio in transverse
direction saturates the KSS bound($\eta/s=1/4\pi$) , it was shown
that the perturbation of the metric with an off-diagonal component
(mixing the anisotropy direction with one of the isotropic
directions) leads to a smaller ratio, $\eta/s<1/4\pi$ and it is a
decreasing function of $a/T$
\cite{Rebhan:2011vd,Mamo:2012sy}.\footnote{As in
\cite{Mateos:2011ix} we identify the temperature and the anisotropy
parameter of the plasma by  $"T"$ and $"a"$, respectively. } This is
the first example of violation of KSS bound not involving higher
derivative theories of gravity \cite{Brigante:2008gz}. Also it was
 shown that the DC conductivity along (transverse) to the anisotropy
direction is a decreasing (increasing) function of the ratio of the
anisotropy parameter to the temperature \cite{Rebhan:2011vd}.

Motivated by these considerations, in this paper we study the effect
of anisotropy on the energy loss of heavy quark propagating through
anisotropic medium. We consider the quark rotating uniformly with
constant velocity in the $xy$ plane. Two possible mechanisms of the
energy loss are considered: drag force and radiation. We compare the
energy loss in an anisotropic plasma to the isotropic case. For
completeness, we will also compare the results with the drag force
and
 jet quenching parameter found in an anisotropic plasma in the last section.

The article is organized as follows. In the next section, we
introduce an anisotropic gravity solution which is a well-defined
IIB supergravity dual to spatially anisotropic $\mathcal{N}=4$ SYM
plasma at finite temperature. Then we will give a discussion about
rotating string in a general background  in section \ref{rsigb}. For
a background with SO(2) symmetry, we find the most general form of
the equations of motion and the critical points of the rotating
string in isotropic plane . This section will be very useful from
technical point of view for our calculations. In section
\ref{rsiscap}, we solve numerically the equations of motion of a
rotating string in the anisotropic plasma (introduced in section
\ref{tgb}) and find the energy loss of a heavy quark. Some analytic
limits will be studied in section \ref{al} and comparison with
linear drag and radiation will be done in section \ref{cwldasr}.
Finally,  we will discuss our results and some possible extensions
for future work in the section \ref{sad}. To complete the
investigation of sections \ref{tgb} and \ref{rsiscap}, appendix
\ref{riap} is devoted to discussion of a rotating quark in an
anisotropic plane. In appendix \ref{ldigb} we give a brief general
review of the drag force which is very useful for comparing our
results in a special limit studied in section \ref{al}.

%---------------------------------------------------------------------------------------------------------------------------------------------------
\section{The gravity background}\label{tgb}

In this section we review the anisotropic background introduced in
\cite{Mateos:2011ix} which is the gravity dual to a deformation of
$\mathcal{N}=4$ SYM adding a $\theta$ term to the action. The
geometry of this solution has two parts: a deformed AdS$_5$, which
has spatially anisotropy but it is asymptotically AdS$_5$, and  an
S$^5$ part. The deformation parameter, $\theta$,  depends on the
anisotropic spatial direction $z$ as $ \theta = 2\pi\,n_{D7}\,z$,
where $n_{D7}$ can be thought as the density of D7-branes
homogeneously distributed along the anisotropic direction. The
D7-branes wrap the S$^5$  and two spacial directions of the AdS$_5$
($xy$-plane).

\begin{figure}[ht]
\centerline{\includegraphics[width=3in]{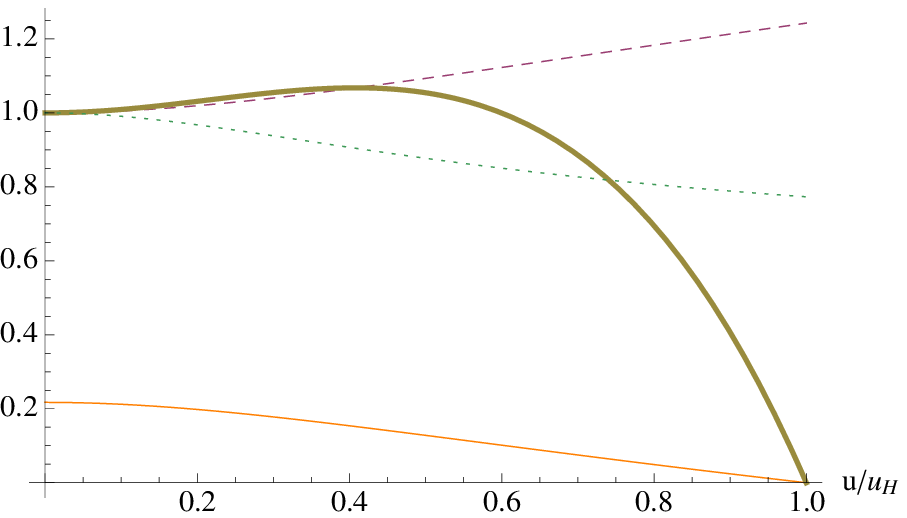}\includegraphics[width=3in]{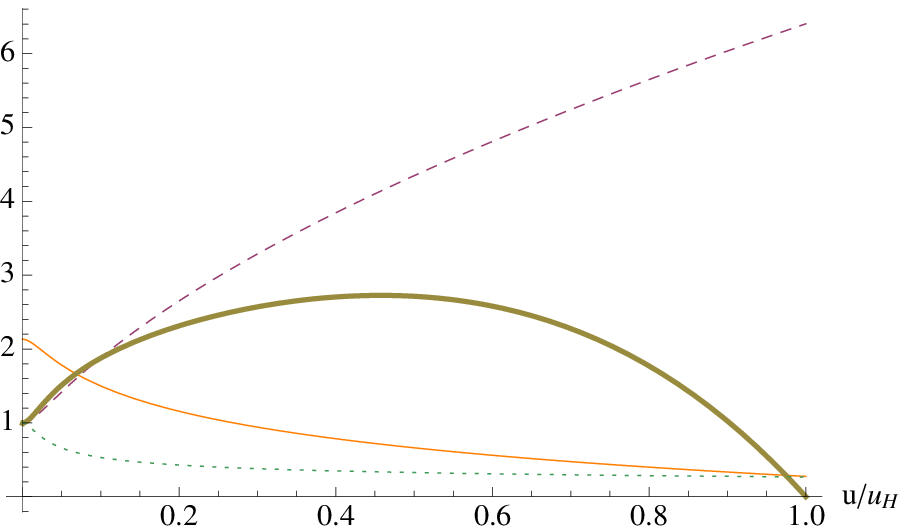}}%
\caption{Metric functions $\tilde{\phi}$ (orange), $\mathcal{H}$ (dashed),
$\mathcal{F}$ (thick), $\mathcal{B}$ (dotted) curves for two different
parameters. Left: $a/T \simeq 4.4$, Right: $a/T\simeq 86$.}\label{background-plots}
\end{figure}

The type IIB supergravity solution in the string frame is given
by \cite{Mateos:2011ix}\footnote{This is the finite-temperature generalization of the
type IIB supergravity solution of \cite{Azeyanagi:2009pr}, the drag
force calculation in this case has been studied in
\cite{Fadafan:2009an}.}
\bea
&&ds^2=\frac{L^2}{u^2}\left(-\mathcal{F}\mathcal{B}dt^2+dx^2+dy^2+\mathcal{H}dz^2+\frac{du^2}{\mathcal{F}}\right)+L^2 e^{\phi/2}ds^2_{S^5}\nn\\
&&\phi=\phi(u),\hspace{10mm}\chi=a\,z,\label{aniso-background}
\eea
where $\phi$, $\chi$, are the dilaton and axion
fields respectively, and  the anisotropy parameter is  $a=\frac{g_{YM}^2n_{D7}}{4\pi}$.
The metric functions are $\mathcal{F}, \mathcal{B}$ and $\mathcal{H}$ which depend
on the holographic radial coordinate, $u$, the radius of horizon, $u_H$, and the anisotropy parameter, $a$.
We choose the range $0<u<u_H$ for the holographic radial coordinate and the boundary is at $u=0$.
$\mathcal{F}$ is the blackening function, $\mathcal{F}(u_H)=0$, and the solution is
asymptotically AdS$_5$ where $\mathcal{F}_b=\mathcal{B}_b=\mathcal{H}_b=1$.
The solutions of the equations of motion are specified by two parameters, the horizon radius,
$u_H$, and the value of dilaton at the horizon, $\phi_h$.
From the boundary theory point of view these parameters map to the temperature,
$T$, and the anisotropy parameter, $a$.
In Fig. \ref{background-plots}, we show two solutions for
different combinations of the parameters: Left $\phi_h\simeq-0.22$ and
$u_H=1$ (or $T\simeq0.33$ and $a/T\simeq4.4$ ), Right $\phi_h\simeq-1.86$ and $u_H=1$
(or $T\simeq0.48$ and $a/T\simeq86$ ).

The metric functions are parameterized by two parameters, the
dilaton field value at the horizon $\phi_h$ and $u_H$ or equivalently by $T$ and $a/T$.
At small $a/T$, the entropy density scales as
in the isotropic case $S_{iso}=\frac{\pi^2}{2}N_c^2 T^3$,
while at large $a/T$ it scales as $S_{aniso} = c_{ent}N^2_c a^{1/3} T^{8/3}$, where
$c_{ent}$ is a constant.
To compare the results in the anisotropic
plasma to that in the isotropic plasma, we consider two
cases: the two plasmas can be taken to have the same
temperatures but different entropy densities, or the same entropy
densities but different temperatures.
Each case  leads to different results.

The field theory coordinates at the boundary are $(t,x,y,z)$ and
there is an SO(2) symmetry in the $xy$ plane. We will consider our
rotating test quark in this plane which can be interpreted as
transverse plane to the beam. We find that, although the quark is
moving in this plane, anisotropy has main effect on the transport
properties. Appendix A is devoted to discussion of a rotating quark
in an anisotropic plane.

One of the main feature of this anisotropic solution is the presence
of a conformal anomaly \cite{Mateos:2011ix} which leads to the fact
that, the thermodynamics of the solution depend on the parameters
$T$ and $a$ separately. This is the cause of broken diffeomorphism
invariance in the holographic direction $u$. We will see that this
is not the case for the rate of energy loss for a rotating string in
this background which takes the form
$\left(dE/dt\right)_{aniso}=\left(dE/dt\right)_{iso}\,
\hat{f}(a/T)$. This kind of behavior is similar to the results of
drag force and jet quenching parameter studied in
\cite{Chernicoff:2012iq,Giataganas:2012zy,Chernicoff:2012gu}.

In next section, we will find the equations of motion of a rotating
string and we will discuss an approach to solve these equations as
generally as possible. Then we will focus on the anisotropic
solution (\ref{aniso-background}) and by comparing the results with
the  isotropic AdS$_5 \times$S$^5$ background, we investigate the
effect of anisotropy on the energy loss of a rotating quark.

%---------------------------------------------------------------------------------------------------------------------------------------------------

\section{Rotating string in a general background}\label{rsigb}

In this section, we will discuss technical details of studying rotating
strings in a general background with SO(2) symmetry. We will use the results of this
section during our calculations in the next sections.
The general form of this static metric is given by
\bea
ds^2=G_{tt}dt^2+G_{\rho\rho}d\rho^2+G_{\psi\psi}d\psi^2+G_{zz}dz^2+G_{uu}du^2,\label{general-bg}
\eea
In our coordinates, $u$ denotes the holographic radial coordinate of the black
hole geometry and $t, \rho, \psi$ and $z$ label the directions along the
boundary at $u=0$.

In the gravity dual, a heavy test quark corresponds to an open
string with one endpoint on a D3-brane at $u=0$. The string hangs
down into the bulk, extending towards the black-hole horizon at $u=
u_H$. In order to study the energy loss for a quark moving on a
circle, we must first find the world-sheet of the string spiraling
downward from the quark moving on a circle at $u=0$ and then
calculate the energy flowing down the string. We choose to
parameterize the two-dimensional world-sheet of the rotating string
according to
\be
X^{\mu}(\tau,\sigma)=\left(t=\tau,\,u=\sigma,\,\psi=\omega\,t+\theta(\sigma),\,\rho=\rho(\sigma),\,z=0\right),\label{ansatz}
\ee
where $\mu$ runs over five dimensions of the bulk theory.

The Nambu-Goto action is
\be
S=-\frac{1}{2\pi\alpha'}\int d\tau d\sigma \mathcal{L}=-\frac{1}{2\pi\alpha'}\int d\tau d\sigma
\sqrt{-\det g_{\mu\nu}},
\ee
where the world-sheet metric $g_{\mu\nu}$ is the
induced metric from background, $g_{\mu\nu}=G_{MN}\rd_\mu X^M\rd_\nu X^N$ in which $G_{MN}$ is the five dimensional background.
Using the general form of the  (\ref{general-bg}) and
the rotating string's ansatz (\ref{ansatz}), one can show that the
Nambu-Goto Lagrangian density is given by
\bea
\mathcal{L}=\left[-G_{uu}(G_{tt}+G_{\psi\psi}\omega^2)-G_{\rho\rho}(G_{tt}+G_{\psi\psi}\omega^2){\rho'}^2-G_{tt}G_{\psi\psi}{\theta'}^2\right]^{1/2}.\label{lagrangian}
\eea
Here, the prime denotes a derivative with respect to the
holographic radial coordinate $u$.

One important feature of this Lagrangian density is $\theta$ dose not appear explicitly in the action and so the momentum conjugate to this field is constant. As we discus in appendix \ref{riap}, this is not the case for a general rotation in anisotropic background.
Then the equation of motion for $\theta$ is given by
\be
\Pi=\frac{\rd \mathcal{L}}{\rd
\theta'}=-\frac{G_{tt}G_{\psi\psi}\theta'}{\mathcal{L}}=\text{const.},\label{eom-theta}
\ee
or equivalently by
\be
{\theta'}^2=-\Pi^2\frac{(G_{tt}+G_{\psi\psi}\omega^2)(G_{uu}+G_{\rho\rho}{\rho'}^2)}{G_{tt}G_{\psi\psi}(G_{tt}G_{\psi\psi}+\Pi^2)},\label{eom-theta1}
\ee
where $\Pi$ is the momentum conjugate to the angular coordinate
$\psi$. The LHS of the above equation is positive definite so, the RHS
should behave in similar manner.
However, the string  stretched from the boundary to the horizon and
the $(G_{tt}+G_{\psi\psi}\omega^2)$ term changes the sign in between these points.
So, the RHS has a critical point
where both numerator and denominator change the
sign while the overall sign doesn't change. This critical
point is called as $u_c$ and is defined by
\bea
&&G_{tt}(u_c)+G_{\psi\psi}(u_c)\omega^2=0,\label{critical1}\\
&&G_{tt}(u_c)G_{\psi\psi}(u_c)+\Pi^2=0.\label{critical2} \eea which
simplifies to \bea
|G_{tt}(u_c)|=\Pi\,\omega,\hspace{20mm}|G_{\psi\psi}(u_c)|=\frac{\Pi}{\omega}.\label{critical}
\eea

For the usual case of an isotropic background which was studied in \cite{Fadafan:2008bq}, the geometry is given by
\be
ds^2=\frac{1}{u^2}\left(-(1-\pi^4T^4u^4)dt^2+d\rho^2+\rho^2d\psi^2+dz^2+\frac{u^4\,du^2}{(1-\pi^4T^4u^4)}\right),
\ee
and  \eqref{critical1} and \eqref{critical2}  reduce to
\bea
&&u_c=\frac{1}{\pi^2T^2}\sqrt{-\frac{\Pi\omega}{2}+\frac{1}{2}\sqrt{\Pi^2\omega^2+4\pi^4T^4}},
\label{critical-iso}\\
&&\rho_c=u_c\sqrt{\frac{\Pi}{\omega}},
\eea
which are the same as the critical values in
\cite{Fadafan:2008bq}.

We use the following form of the metric components,
\be
ds^2=f(u)\left(g(u)dt^2+d\rho^2+\rho(u)^2d\psi^2+h(u)du^2\right)+G_{zz}(u)dz^2,\label{general-bg1}
\ee
so that the equation of motion for $\rho$ field can be written in a more compact form.
Using this form of the metric (\ref{general-bg1}) and the critical equations (\ref{critical}), one can find the radius of the rotation at the critical point as
\be
\rho_c=\frac{1}{f_c}\sqrt{\frac{\Pi}{\omega}},\label{rho-c-simple}
\ee
where $f_c=f(u_c)$.
Using the equation of motion for $\theta$  (\ref{eom-theta}),
one can show the equation of motion for $\rho(u)$ is given by
\bea
&&2 g^2 h^2 \left(\Pi ^2-\omega^2 f^2 \rho^4\right)\nn\\
&&-\left[h \rho^3 \left(-2 f g^2 \left(g+\omega^2 \rho^2\right) f'+\left(\Pi^2 \omega^2-f^2 g^2\right) g'\right)+g \rho \left(g+\omega^2 \rho^2\right) \left(\Pi^2+f^2 g \rho^2\right) h'\right] \rho'\nn\\
&&+2 g^2 h \left(\Pi^2-\omega^2 f^2 \rho^4\right) \rho'^2\nn\\
&&+\rho^3 \left[2 f g^2 \left(g+\omega^2 \rho^2\right) f'+\left(-\Pi ^2 \omega^2+f^2 g^2\right) g'\right] \rho '^3\nn\\
&&+2 g h \rho \left(g+\omega^2 \rho^2\right) \left(\Pi ^2+f^2 g
\rho^2\right) \rho''=0.\label{eom-rho}
\eea
To find the radius of rotation at the boundry one needs to solve this equation numerically, even for simpleset cases.

The rotating-string solution starts at $(u=0, \rho =l)$ passes
through the critical point $(u_c,\rho_c)$ and approaches the black
hole horizon located at $u =u_H$ and a finite value of $\rho$ that
is greater than $\rho_c$, which in turn is greater than $l$. By
expanding \eqref{eom-rho} around $u_c$, one can find a series
expansion of $\rho(u)$ as
$\rho(u)=\rho(u_c)+(u-u_c)\rho'(u_c)+\frac{1}{2}(u-u_c)^2\rho''(u_c)+...\,$.
For example, the first term in expansion  $\rho'(u_c)$ has four
solutions \bea
&&\rho'^{(1,2)}_c=\frac{1}{\left(4 f_c g_{c}^3 \rho_c f'_c\right)}\bigg[4 f_c^2 g_c^3 h_c+g_c^3 \rho_c^2 f_c'^2+\omega^2 g_c^2 \rho_c^4 f_c'^2+2 f_c g_c^2 \rho_c^2 f_c' g_c'-\Pi ^2 {g_c}'^2\nn\\
&&\pm\left(16 f_c^2 g_c^6 h_c \rho_c^2 f_c'^2+\left(4 f_c^2 g_c^3 h_c+g_c^3 \rho_c^2 {f_c}'^2+\omega^2 g_c^2 \rho_c^4 {f_c}'^2+2 f_c g_c^2 \rho_c^2 f'_c g_c'-\Pi ^2 {g_c'}^2\right)^2\right)^{1/2}\bigg],\nn\\
&&\rho'^{(3,4)}_c=\pm\sqrt{-h_c},
\eea
Obviously, the 3rd and 4th solutions are not physical. The only
acceptable solution is the 1st solution (positive sign) which leads
to the physical behavior, $\rho(u=0)=l<\rho_c$.\footnote{In most of the case $f(u)\propto u^n$ which leads to a positive (negative) solution of $\rho'_c$ for $n>0$ ($n<0$).}
In principle, once we find the series expansion of $\rho(u)$
around $u_c$, it would be possible to guess the closed form of
$\rho(u)$. In general it is not easy to find $\rho(u)$ in this way
and the only known result is the simple case of rotating string at
zero temperature $\mathcal{N}=4$ SYM plasma studied in
\cite{Athanasiou:2010pv}.\footnote{K.B.F thanks D.Nickel for
discussion on this point.}

We use numerical methods to find the shape of the rotating string.
Given angular momentum $\omega$ and constant of motion $\Pi$, one
finds $(u_c, \rho_c)$ point from \eqref{critical} and
\eqref{rho-c-simple}. But there is a problem at this point, the
equation of motion (\ref{eom-rho}) (and the Lagrangian
(\ref{lagrangian})) is singular at the critical point. So, by
starting from this point, one can not find the numerical solution
for the equation of motion of $\rho$   (\ref{eom-rho}).\footnote{
This challenge have been seen also in some other calculation
\cite{Chesler:2008uy} in a range of $u$ and, it leads to using the
Polyakof action instead of Nambu-Goto action.} Fortunately this
problem appears only at one point and we can use the equation of
motion (\ref{eom-rho}) and expand it up to any order that we want.
Since the equation (\ref{eom-rho}) is a second order differential
equation, we must know two boundary conditions to find the solution
uniquely. We can find the second derivative of the $\rho(u)$ at
$u_c$, $\rho''(u_c)$ and then solve the equation numerically by
expanding from $u_c-\epsilon$ ($u_c+\epsilon$) to the boundary
(horizon).

In general, the energy loss rate of the world-sheet is given by
\be
\frac{dE}{dt}=\Pi^\sigma_t=-\frac{\delta S}{\delta (\rd_\sigma
X^0)},
\ee
In our case, by using equations (\ref{lagrangian}),
(\ref{eom-theta}) and (\ref{critical}), one can show that there are
some different expansions for the rate of energy loss which are very useful,
\be
\frac{dE}{dt}=-\frac{G_{tt}G_{\phi\phi}\omega \theta'}{2 \pi \alpha'
\mathcal{L}}=\frac{\Pi \omega}{2\pi
\alpha'}=\frac{|G_{tt}(u_c)|}{2\pi \alpha'}.\label{energy-loss}
\ee
In appendix \ref{ldigb} we will show that the general formula for the rate of
energy loss for a string moving on a line is the same as the last
equality in the above equation.
The difference is a different value for the critical point of the radial holographic coordinate, $u_c$.

To find and investigate the solution in terms of the parameters of
the boundary theory, we will focus on the strongly-coupled
anisotropic plasma in the next section and compare our results with
the rotating string in the isotropic background which is the
AdS$_5\times$S$^5$ theory with $\mathcal{N}=4$ SYM theory on the
boundary. Then, we will compare the energy loss of a rotating string
to the drag force in the strongly-coupled anisotropic plasma studied
in \cite{Chernicoff:2012iq, Giataganas:2012zy} and the radiation
results on isotropic case investigated in \cite{Fadafan:2008bq}.
%---------------------------------------------------------------------------------------------------------------------------------------------------
\section{Rotating string in strongly-coupled anisotropic plasma}\label{rsiscap}

To study the energy loss of the rotating quark in strongly-coupled
anisotropic plasma, one should solve equation (\ref{eom-rho}) to
find the radius function $\rho(u)$. The general behavior of the
angular function, $\theta(u)$, is the same as in the isotropic case
studied in \cite{Fadafan:2008bq}, so we are not going to find the
angle function $\theta(u)$, which has no effect on energy lost.

Valuable results are the comparison of the anisotropic case with
the isotropic background at the same temperature or at the same
entropy density.
We do these comparisons both for the gravity
side (bulk) and the filed theory side (boundary) of the AdS/CFT
correspondence.

\begin{figure}[ht]
\subfigure[$a/T \simeq
4.4$]{\includegraphics[width=2in]{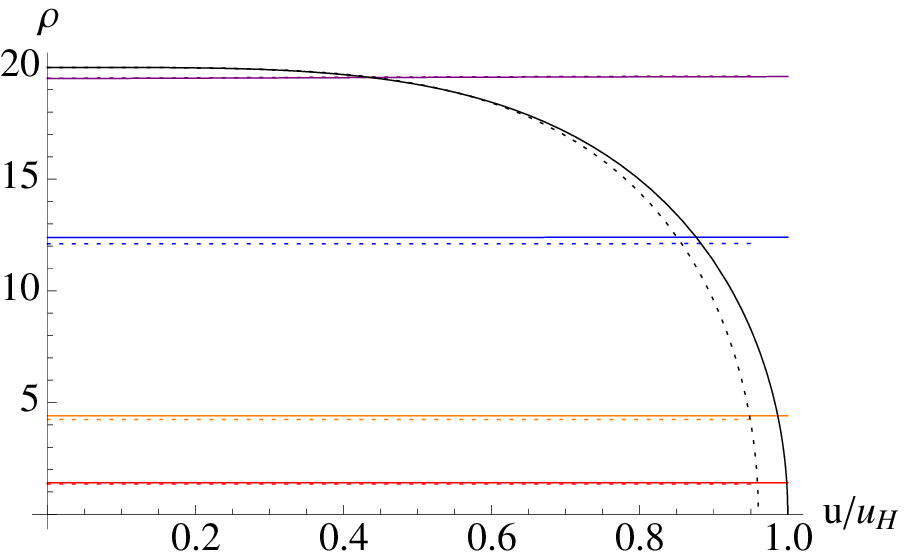}
\includegraphics[width=2in]{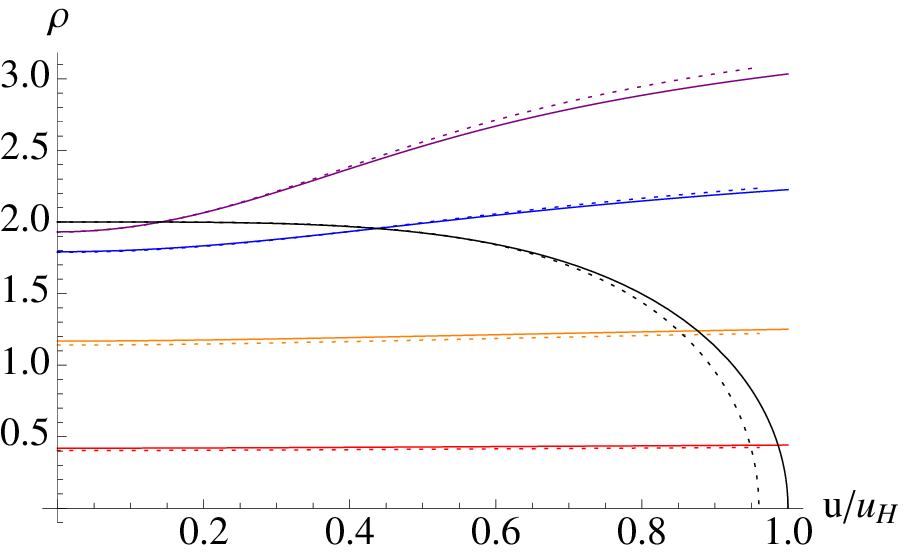}\includegraphics[width=2in]{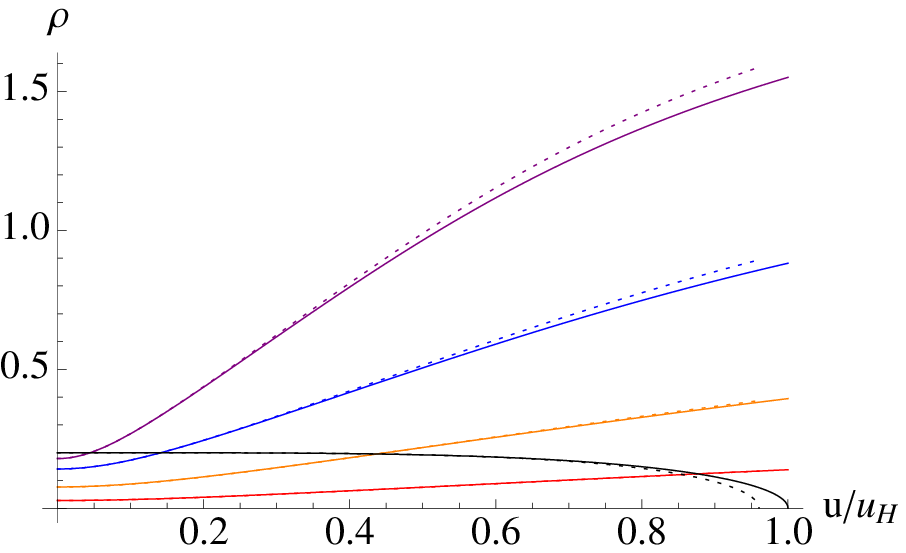}}
\subfigure[$a/T \simeq
86$]{\includegraphics[width=2in]{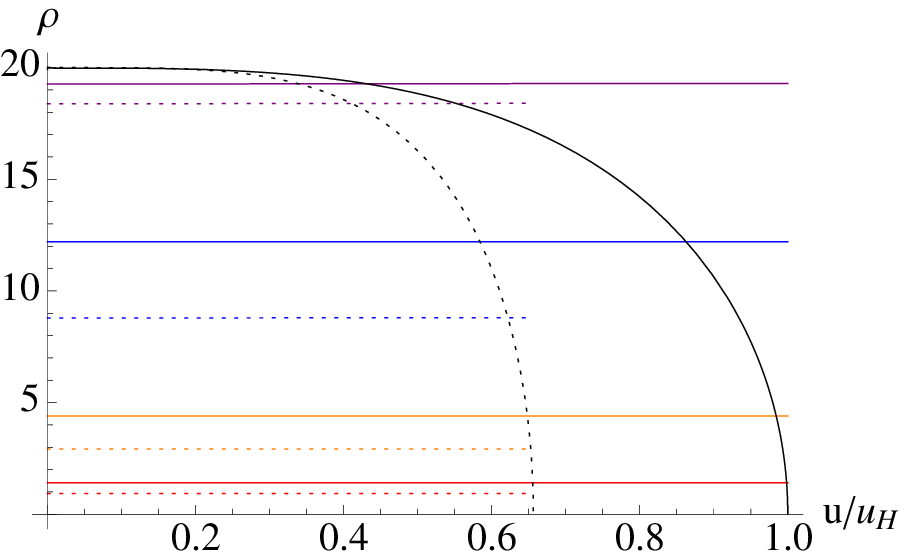}\includegraphics[width=2in]{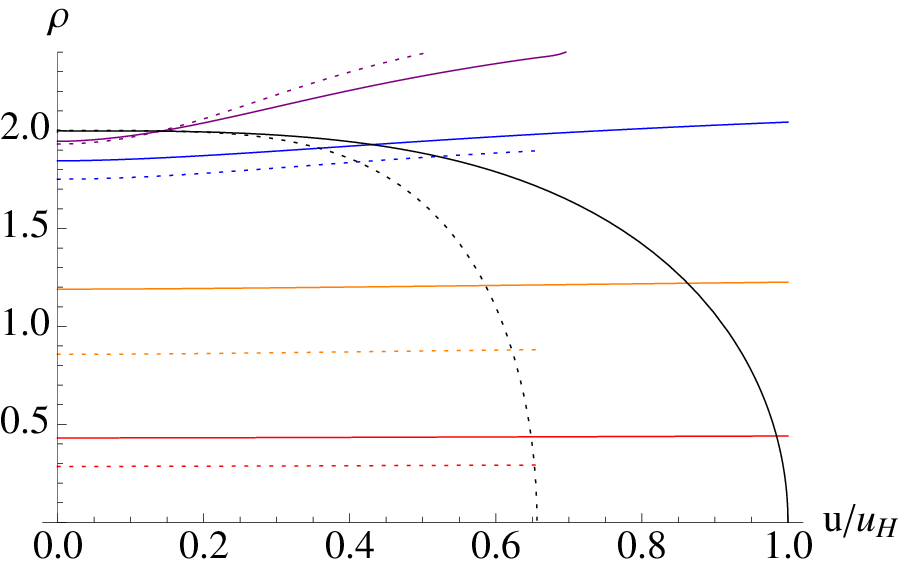}\includegraphics[width=2in]{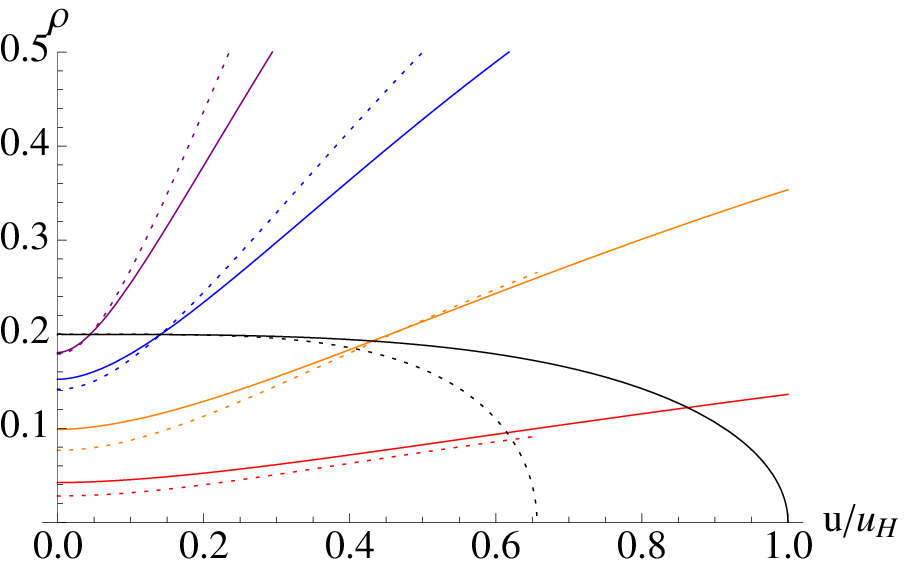}}
\caption{The radial dependence $\rho(u)$ of spiraling string hanging
down from rotating quarks in anisotropic (solid curves) and isotropic (dotted curves) media for various different choices of
parameters: $\omega= 0.05, \omega= 0.5$ and $\omega= 5.0$ from left
to right, and the set of values for the constant of motion $\Pi=$0.1
(red), 1 (orange), 10 (blue), 100 (purple) for all of the figures.
In (a) and (b), the anisotropic parameter is specified as $a/T
\simeq 4.4$ and $a/T \simeq 86$, respectively. This comparison with
the isotropic plasma has been done at the same temperature.}\label{rho-u-temp}
\end{figure}%

By assumption, the temperature and the entropy density of the two sides
of the AdS/CFT correspondence are the same. Therefore it's enough  to find
these parameters for the gravity side by using simple definitions of
the temperature and entropy density. We will study the energy loss of a rotating quark
in an anisotropic plasma by comparing the results with the isotropic case both with the same temperature
and with the same entropy.

Plugging the anisotropic background (\ref{aniso-background}) into the
critical equation \eqref{critical}, the equation of motion for $\rho(u)$
(\ref{eom-rho}), and the equation which determines the $\rho_c'^{}$;
one can reduce these equations to a simpler form. As we mentioned in
section \ref{rsigb}, to find  numerical results for $\rho(u)$ we should
integrate from $u_c-\epsilon$ ($u_c+\epsilon$) to the boundary
(horizon).

In Fig.\ref{rho-u-temp}, we compare the anisotropic plasma (solid
curves) with isotropic plasma (dotted curves) in the same
temperature. These plots show the radial dependence $\rho(u)$ of a
spiraling string hanging down from rotating quarks for different
angular velocities: $\omega=0.05, 0.5, 5.0$, from left to right. In
this figure the constant of motion (angular momenta) is given by
$\Pi=$0.1 (red), 1 (orange), 10 (blue), 100 (purple). The effect of
anisotropy
 is given by the ratio of anisotropic parameter to the temperature $a/T \simeq 4.4$ (a) and $a/T \simeq 86$ (b). Note that the string is not going into the horizon so the dotted (solid) curves finished at the horizon of isotropic (anisotropic) solutions.
\begin{figure}[ht]
\subfigure[$\frac{ N_c^{2/3}a}{s^{1/3}} \simeq
2.5$]{\includegraphics[width=2in]{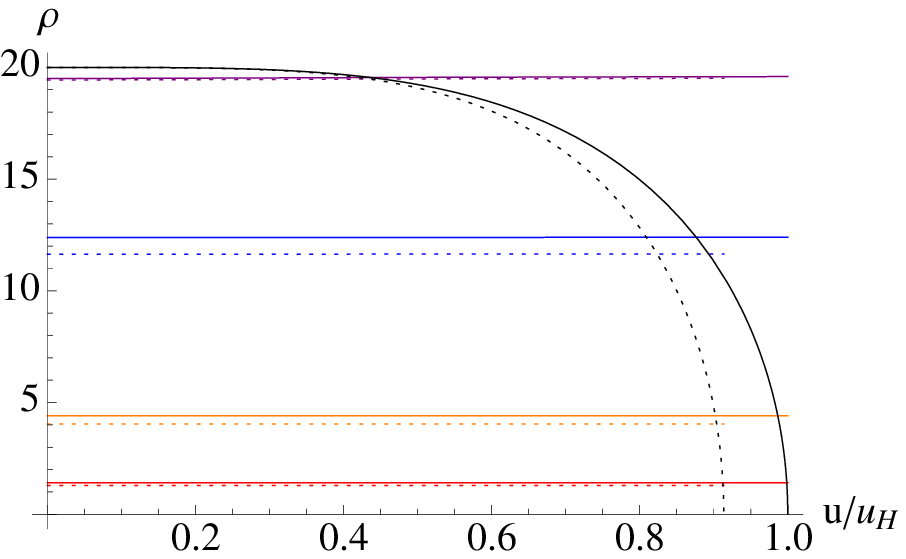}\includegraphics[width=2in]{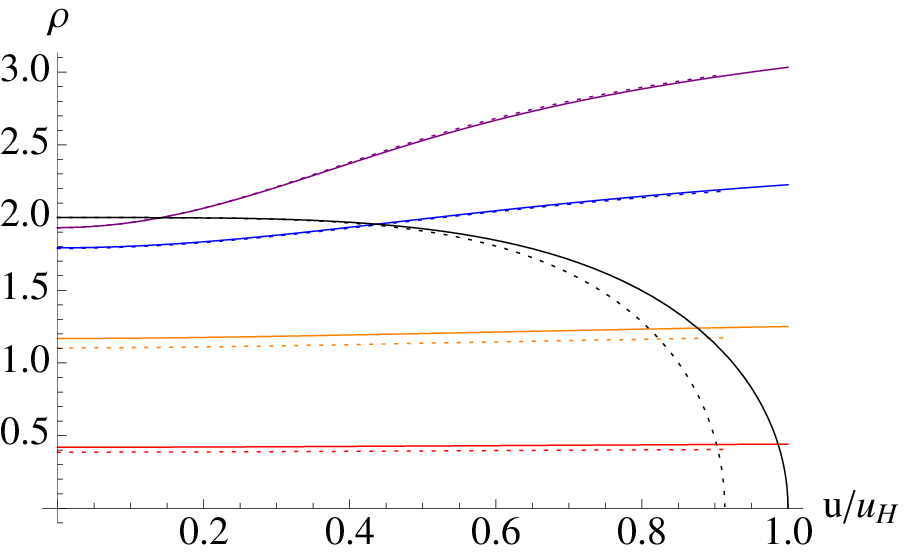}\includegraphics[width=2in]{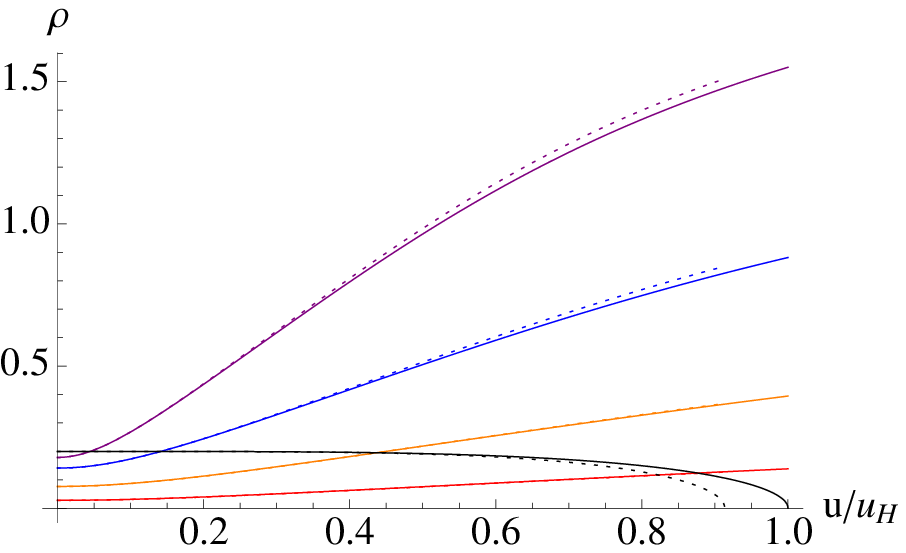}}
\subfigure[$\frac{ N_c^{2/3}a}{s^{1/3}} \simeq
35.5$]{\includegraphics[width=2in]{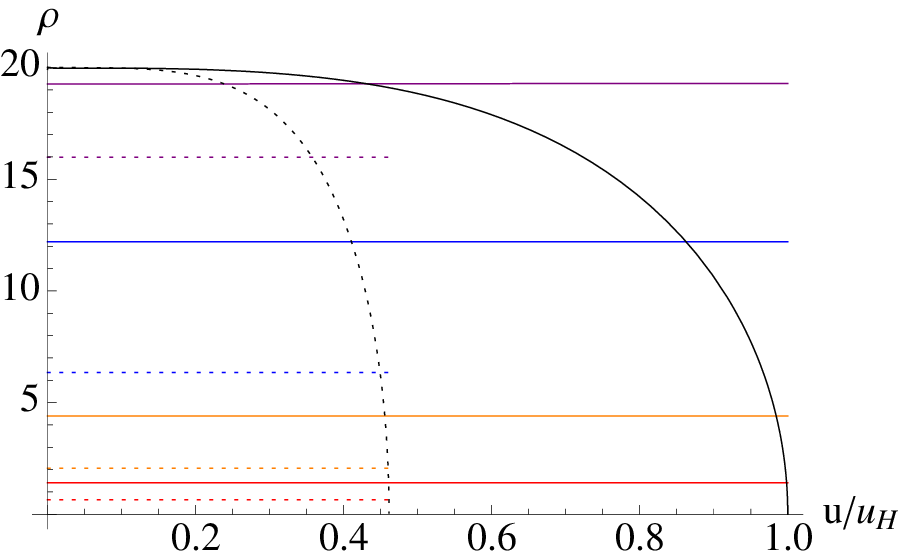}\includegraphics[width=2in]{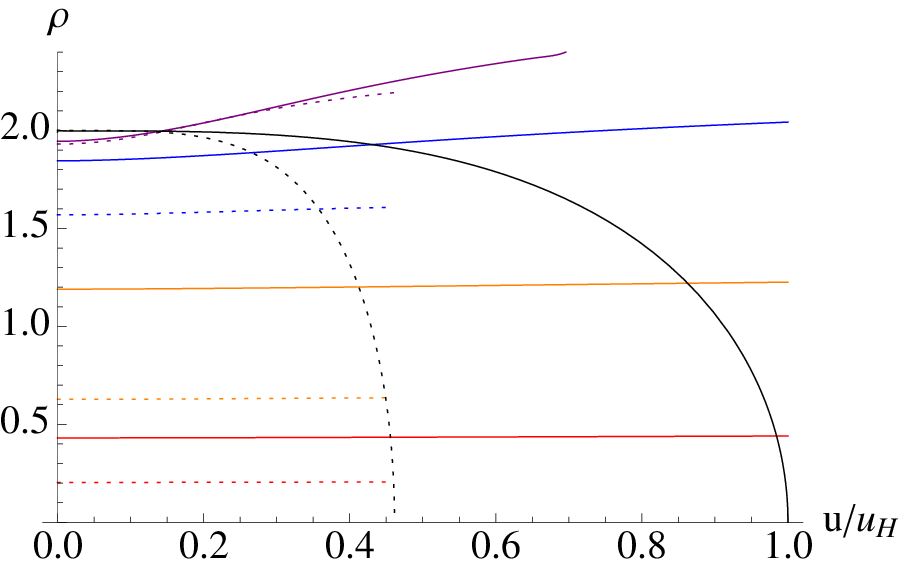}\includegraphics[width=2in]{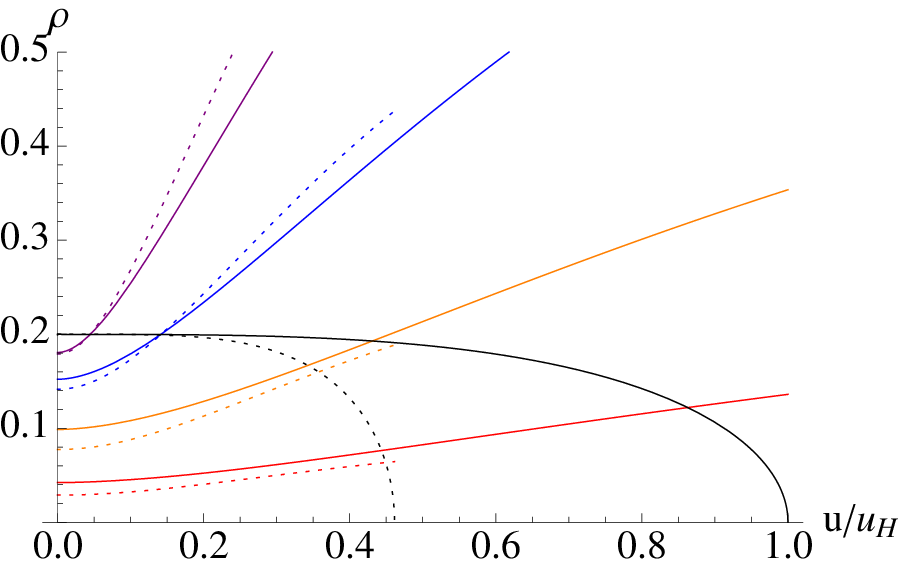}}
\caption{The radial dependence $\rho(u)$ of spiraling string hanging
down from rotating quarks in anisotropic (solid curves) and isotropic
(dotted curves) media for various different choices of parameters:
$\omega= 0.05, \omega= 0.5$ and $\omega= 5.0$ from left to right,
and the set of values for the constant of motion $\Pi=$0.1 (red), 1
(orange), 10 (blue), 100 (purple) for all of the figures. In (a) and
(b), the anisotropic parameter is specified as $\frac{
N_c^{2/3}a}{s^{1/3}} \simeq 2.5$ and $ 35.5$, respectively. This
comparison with the isotropic plasma has been done at the same
entropy.}\label{rho-u-entropy}
\end{figure}%

On the other hand, for fixed $\omega$ and $\Pi$, which means the role of energy loss is fixed,
the bigger radius of rotation on the boundary means a larger  velocity for the quark.

We summarize the important results of the plots as follows:
\begin{itemize}
\item{The anisotropy mainly effected the shape of the spiraling string. It is easy to see that by increasing the anisotropy the radius of rotation significantly changes.}
\item{While in most of the cases the radius of rotation on the boundary in anisotropic cases is larger than isotropic cases, for small anisotropy, $a/T\simeq4.4$, and small angular velocity, $\omega=0.05$, this is not the case. }
\item{For the fixed $\omega$ and $\Pi$, the radius on the boundary
is bigger than in the isotropic case which means a larger velocity of the quark in the field theory.
Using the energy loss relation \eqref{energy-loss}, $\frac{dE}{dt}\propto\Pi\omega$,
one can show that for the same velocity $v$ and the same angular velocity $\omega$ the energy loss in the anisotropic plasma is less than the energy loss in the isotropic case
(except for the small $\omega$ and large $v$ which is similar to the drag force \cite{Chernicoff:2012iq}).
We will discuss  this point in the later  sections.}
\item{ For the case of small $\omega$ (0.05), the isotropic and anisotropic curves with the same $\Pi$ do not cross each other and each curve is almost straight.
So the energy loss should be very similar to the drag force and effect of radiation  is very small.
We will show (both numerically and analytically) that this is the case. }
\item{For a bigger $\omega$ (0.5) the curves cross each other for enough large $\Pi's$.
We will see that that this crossing may be assumed to be a measure for interplay between the linear drag channel
and the radiation channel of energy loss. }
\item{In large $\omega$ (5.0), for small $a/T$ the radius on the horizon is
smaller than the isotropic case and there is no crossing. But  for large anisotropy the
curves cross each other and the radius on the horizon is larger than
isotropic case, both for small and large anisotropy.}
\end{itemize}

In Fig.\ref{rho-u-entropy}, we show the results for the same entropy
density. The general behavior of the plots are similar to the plots
we find for the same temperature cases in Fig.\ref{rho-u-temp}. The
main difference is for small $a/T$, there is no strange behavior as
like as the same temperature cases we discussed earlier. So the
radius of rotation on the boundary in anisotropic plasma is smaller
than isotropic case for fixed $\omega$ and $\Pi$ in both small and
large anisotropy. Again, we specified the anisotropic results by
solid curves and the isotropic plasma
 by dotted curves, $\omega=0.05, 0.5,5.0$ from left to right and
$\Pi=$0.1 (red), 1 (orange), 10 (blue), 100 (purple).

Here we compare the rotating string in anisotropic and isotropic
cases with fixed $\Pi$ and $\omega$, which means  the fixed energy
loss rate by using eq.(\ref{energy-loss}). We investigate this
comparison from two different point of views: gravity theory and
boundary theory. The parameters of solutions from gravity point of
view are $\omega$ and $\Pi$ but, from boundary theory they are
$\omega$ and $v=l\omega$. Therefore from gravity point of view we
find the ratio of energy loss in anisotropic and isotropic plasmas
as a functions of $\Pi$, and from  boundary theory point of view we
find this ratio as a functions of $v$.

%------------------------------------------------------------
\subsection{The energy loss}\label{tel}

In this section, we study the energy loss of a heavy test quark.
Plugging the anisotropic background metric (\ref{aniso-background}) into the
general formula for the rate of  energy loss (\ref{energy-loss}) one
can show that
\be
\frac{dE}{dt}=\frac{\Pi\,\omega}{2\pi\alpha'}=\frac{1}{2\pi\alpha'}\frac{\mathcal{F}_c\mathcal{B}_c}{u_c},\label{energy-loss-aniso}
\ee
where $\mathcal{F}_c$ and $\mathcal{B}_c$ are the value of these
functions at the critical point $(u_c, \rho_c)$ which is defined by
\bea
\frac{\mathcal{F}_c\mathcal{B}_c}{u_c^2}=\Pi\omega,\hspace{20mm}\rho_c=u_c\sqrt{\frac{\Pi}{\omega}}\label{critical-aniso}
\eea
Therefore, from bulk point of view, the energy loss just
depends on the value of $u_c$ which is a function of $\Pi$ and
$\omega$ via  equations (\ref{critical-aniso}).

\begin{figure}[ht]
\centerline{\includegraphics[width=4in]{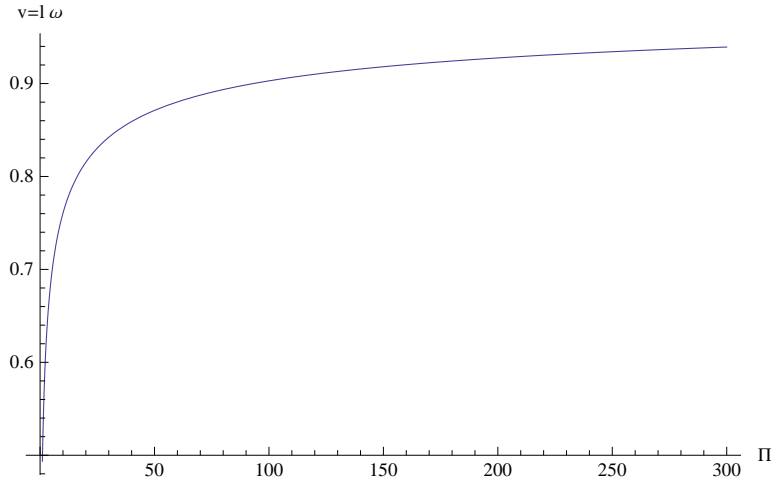}}
\caption{In general for both isotropic and anisotropic solutions the velocity (or equivalently the radius of the rotation) of the end point of the string on the boundary is a monotonically increasing function of $\Pi$.
As an example, in this plot we show the velocity of the quark at the boundary versus the
constant of motion $\Pi$ for $\omega=5.0$ and
$a/T\simeq86$.}\label{Pi-l-general}
\end{figure}

Using \eqref{energy-loss-aniso},  it is straightforward, using
the numerical techniques described before, to evaluate the energy
loss for a quark moving in a circle with angular frequency $\omega$
as a function of  radius $l$ (or equivalently as
a function of velocity of the quark  $v=l\,\omega$).

From the bulk point of view, we pick an $\omega$ and
a series of values of $\Pi$. For each $\Pi$, we obtain the
corresponding string world-sheet.
From that solution, one finds that the
energy loss of the quark increases without bound as $l$ approaches to $1/\omega$,
which is the radius at which the quark would be moving at
the speed of light. We call the radius of the rotating quark in the
anisotropic and isotropic medium $l_{aniso}$ and $l_{iso}$,
respectively.

The radius of the rotation on the boundary (or the velocity of the rotating quark $v=l\omega$) is a monotonic increasing function of $\Pi$ for fixed $\omega$ in both isotropic and anisotropic back grounds.
As an example, we plot the velocity of the quark as a function of $\Pi$ for $\omega=5.0$ and $a/T\simeq86$ in Fig.\ref{Pi-l-general}.
This behavior is general for any value of $\omega$ and $a/T$ and it will be extremely useful in what follows.

\begin{figure}
\subfigure
{\centerline{\includegraphics[width=60mm]{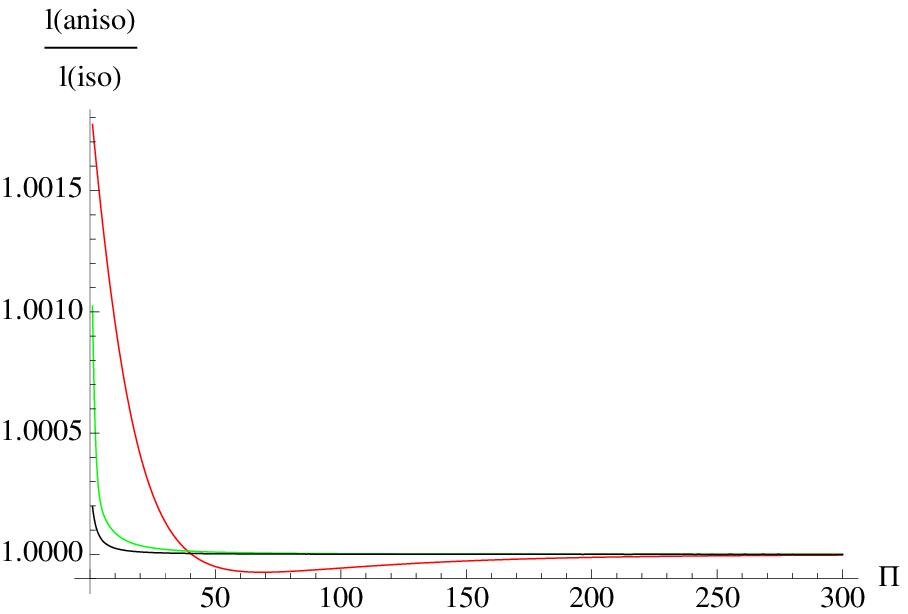}\includegraphics[width=60mm]{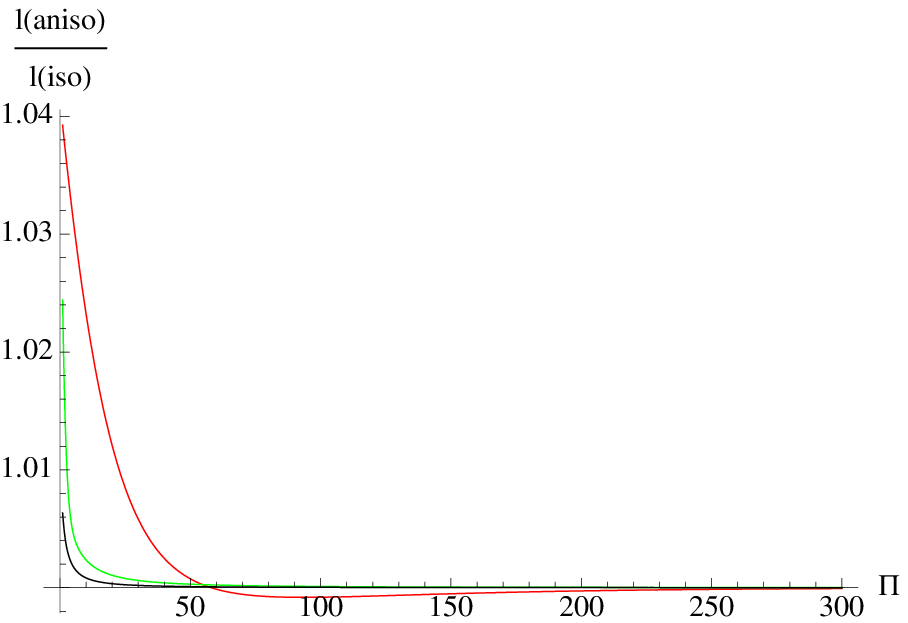}}}
\subfigure
{\centerline{\includegraphics[width=60mm]{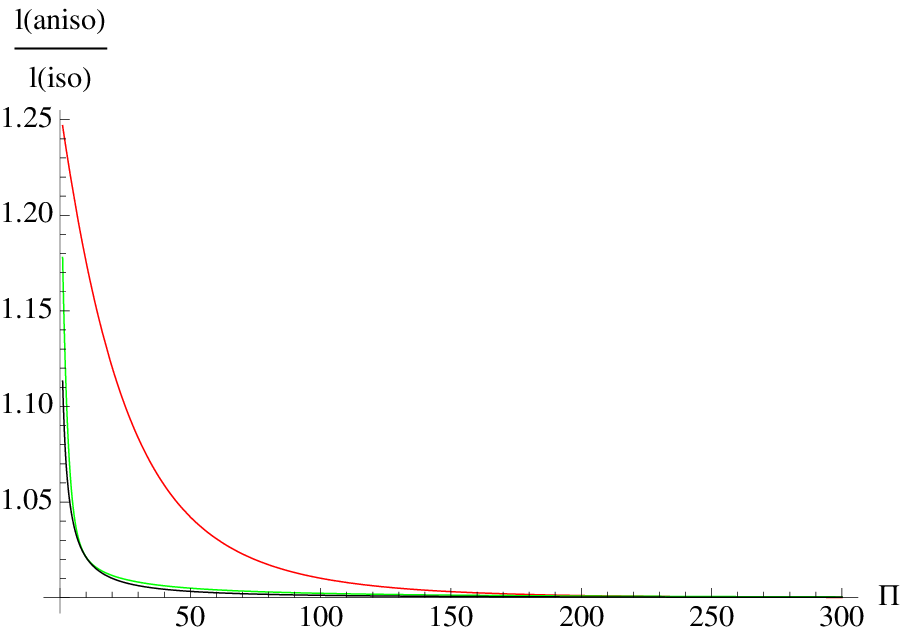}\includegraphics[width=60mm]{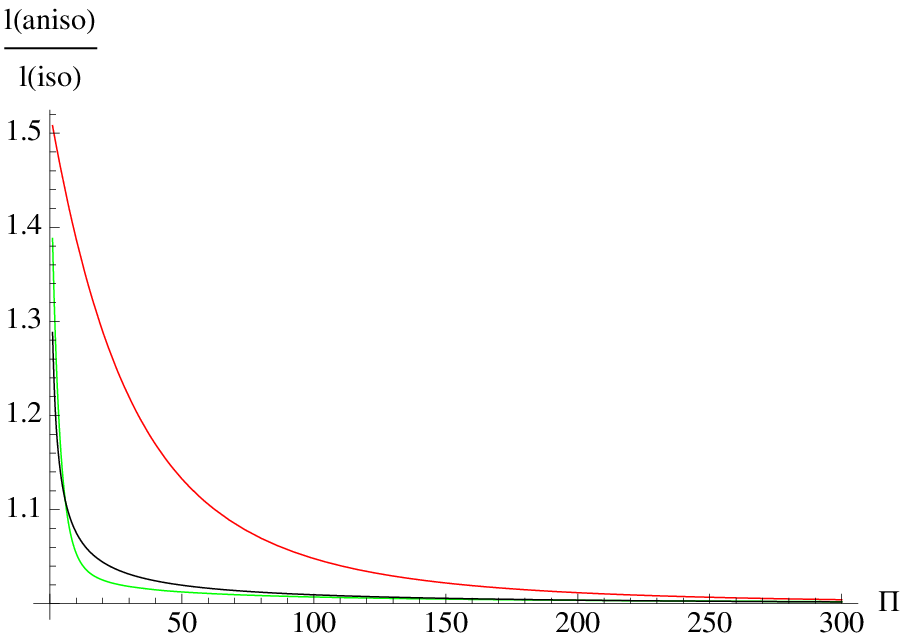}}}
\caption{The ratio of the radius of rotating quark in an anisotropic
medium to the isotropic case as a function of the constant of motion
$\Pi$. This comparison has been done for the same temperature. In
each plot of this figure, the angular velocity is $\omega=$0.05
(red), 0.5 (green), 5.0 (black), and $a/T\simeq$ 0.78 (top-left),
4.4(top-right), 25(down-left),
86(down-right)}\label{laniso-liso-temp}
\end{figure}

\begin{figure}
\subfigure
{\centerline{\includegraphics[width=60mm]{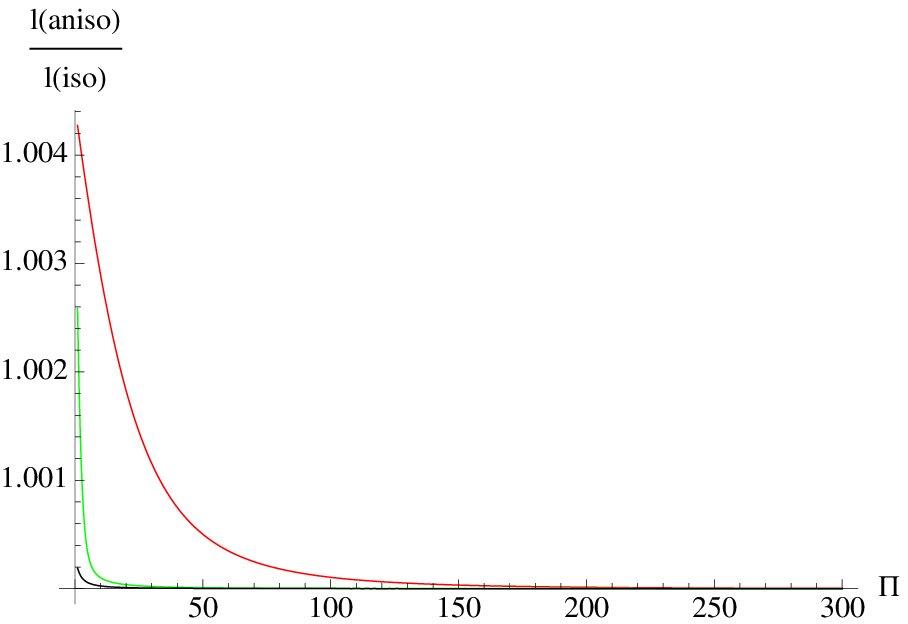}\includegraphics[width=60mm]{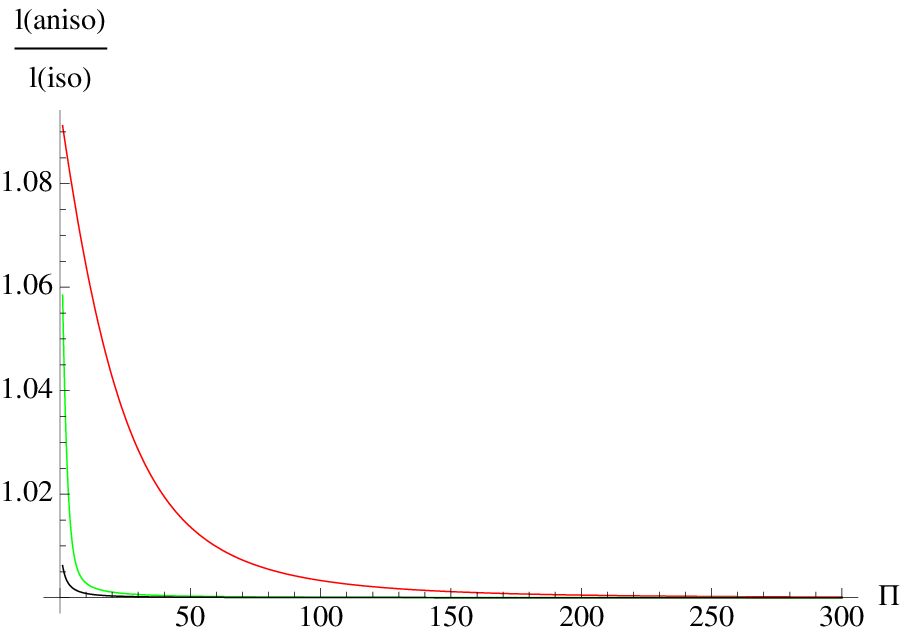}}}
\subfigure{\centerline{\includegraphics[width=60mm]{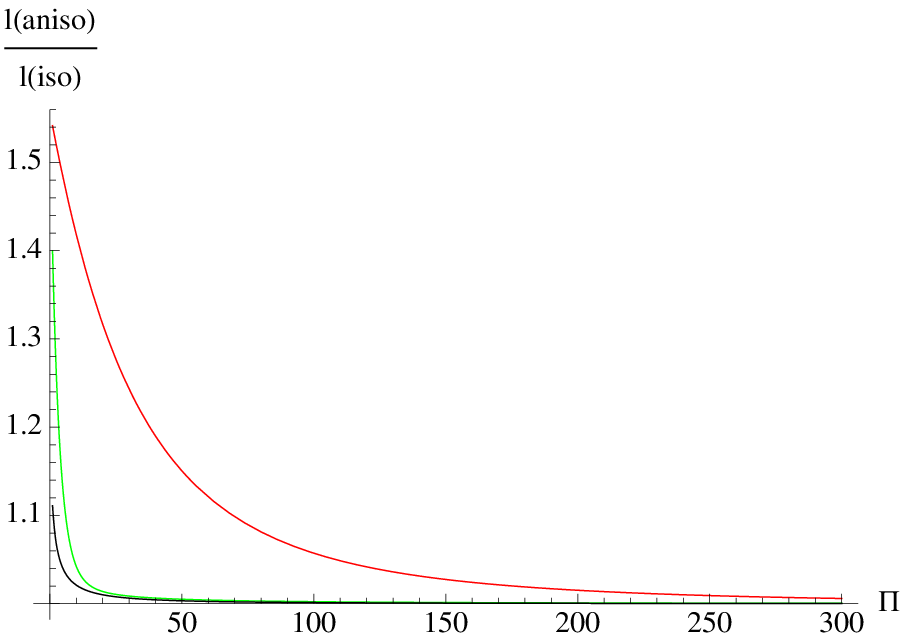}\includegraphics[width=60mm]{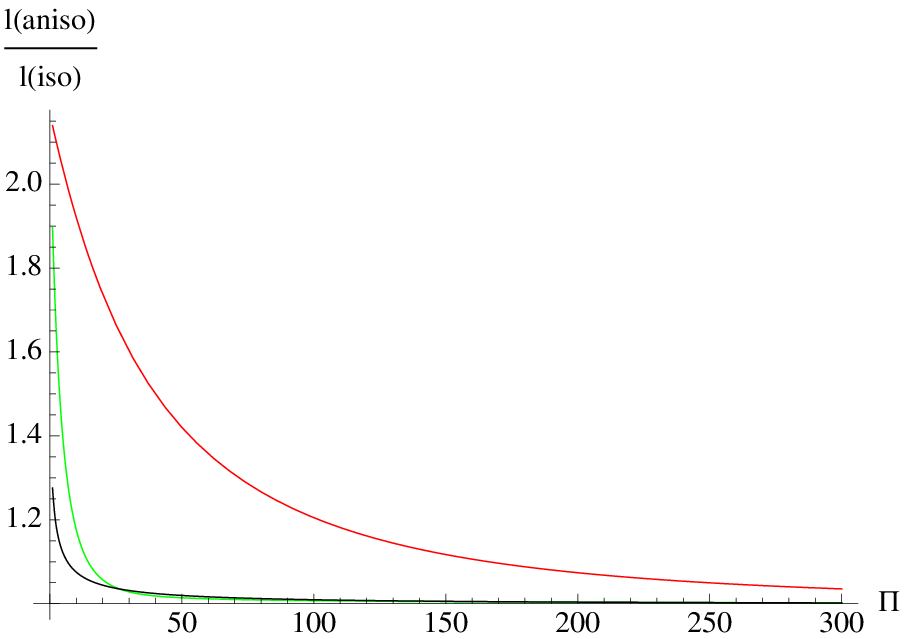}}}
\caption{The ratio of the radius of rotating quark in an anisotropic
medium to the isotropic case as a function of the constant of motion
$\Pi$. This comparison has been done for the same entropy. In each
plot of this figure, angular velocity is $\omega=$0.05 (red), 0.5
(green), 5.0 (black), and $ \frac{ N_c^{2/3}a}{s^{1/3}}\simeq$  0.46
(top-left), 2.5 (top-right), 12(down-left), 35.5
(down-right)}\label{laniso-liso-ent}
\end{figure}

We plot $l_{aniso}/l_{iso}$ (which is as same as $v_{an}/v_{is}$) as a function of
the constant of motion $\Pi$ for different angular velocities, $\omega$ and different $a/T$ for
the same temperature and the same entropy in figures \ref{laniso-liso-temp} and \ref{laniso-liso-ent}, respectively.
Each figure contains  the plots for four anisotropic parameters $a/T\simeq$ 0.78, 4.4, 25, 86 and three angular
velocities $\omega=$0.05 (red), 0.5 (green), 5.0 (black).

There are two general features in all plots:%
\be (1)\hspace{3mm} l_{aniso}>l_{iso}|_{\Pi\rightarrow
0},\hspace{10mm}(2)\hspace{3mm}
l_{aniso}\simeq l_{iso}|_{\Pi\rightarrow \infty}. \ee%
This shows that, for large enough $\Pi$'s the quark can not see the
anisotropy while, the anisotropy has a significant affect on the
rotating quark for smaller $\Pi$'s. We will discuss about this
behavior also from boundary theory point of view.

In Fig.\ref{laniso-liso-temp}, it is clearly seen that for the same
temperature cases, there is a minimum value, which is less than $1$,
for this ratio in small angular velocity, $\omega=0.05$, and small
anisotropy, $a/T$ ($\simeq0.78$ and  $4.4$). This minimum disappear
by increasing the angular velocity and/or $a/T$. But all plots with
the same entropy in  Fig.\ref{laniso-liso-ent} show that $l_{aniso}$ is
bigger than $l_{iso}$ for any $\omega$ and $\Pi$.

According to the first equality in equation
(\ref{energy-loss-aniso}), the energy loss rate is proportional to
$\Pi\,\omega$. On the other hand, in both isotropic and anisotropic
cases the radii of rotation at the boundary are monotonically
increasing with respect to $\Pi$ (e.g. Fig. \ref{Pi-l-general}). So
from the boundary theory point of view, one can conclude that for
$l_{aniso}/l_{iso}>1$ ($l_{aniso}/l_{iso}<1$) the rate of energy
loss in an anisotropic plasma is smaller (larger) than the isotropic
case, for  fixed $v$ and $\omega$. It is easy to see that in these
conditions the rate of energy loss in an anisotropic plasma is less
than isotropic case in general except for the small value of
$\omega(0.05)$ and small $a/T(\simeq0.78)$. In this case for large
enough $\Pi(\gtrsim40)$, corresponding to $v\simeq0.9$,  the energy
loss rate in the anisotropic plasma is more than the isotropic case.
This is in perfect agreement with the drag force results
\cite{Chernicoff:2012iq, Giataganas:2012zy}. We expect that for a
small angular velocity the dominant part of the energy loss is due
to the drag force, and the above agreement bolsters this idea. We
checked this behavior for a range of $a/T$ and we see that this
happens for $a/T\lesssim12$.

\begin{figure}[ht]
\subfigure[The same temperature, $a/T\simeq$1.4 (red), 4.4 (blue), 12 (green), 25 (brown).]{\includegraphics[width=2in]{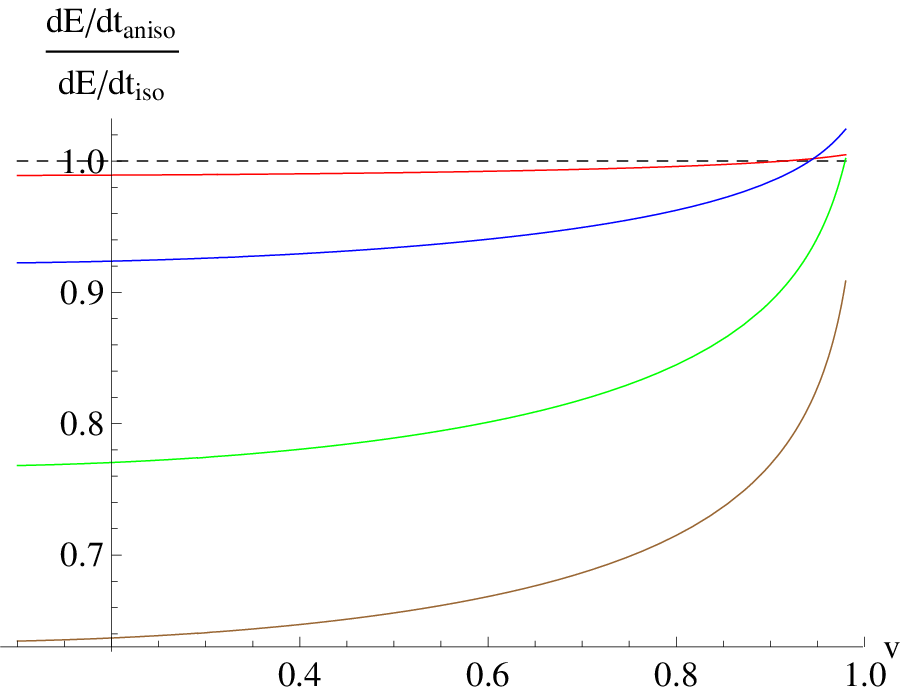}\includegraphics[width=2in]{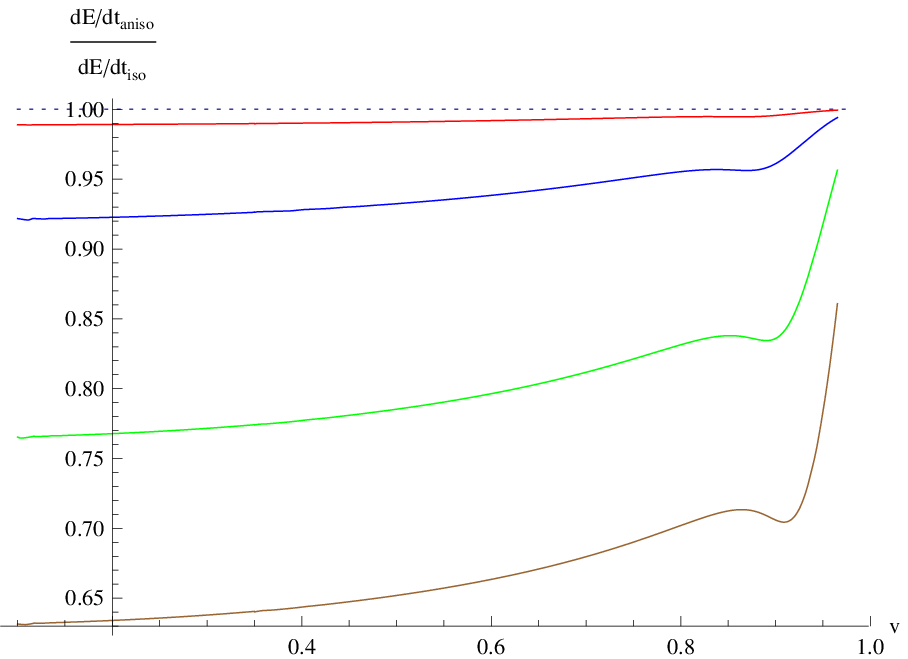}\includegraphics[width=2in]{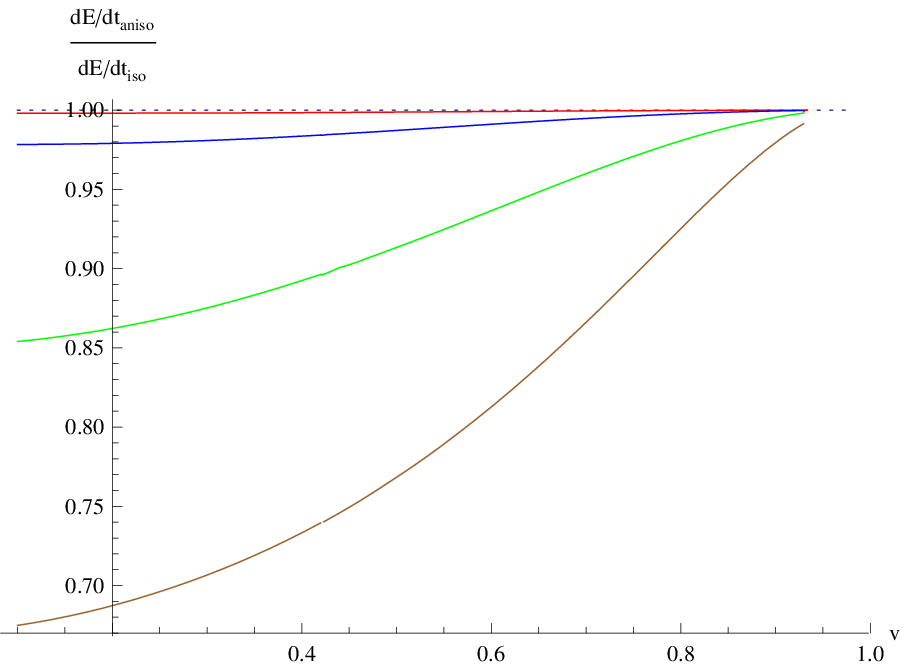}}
\subfigure[The same entropy, $N_c^{3/2}a/s^{1/3}\simeq$0.80 (red), 2.5 (blue), 6.2 (green),
 12 (brown).]{\includegraphics[width=2in]{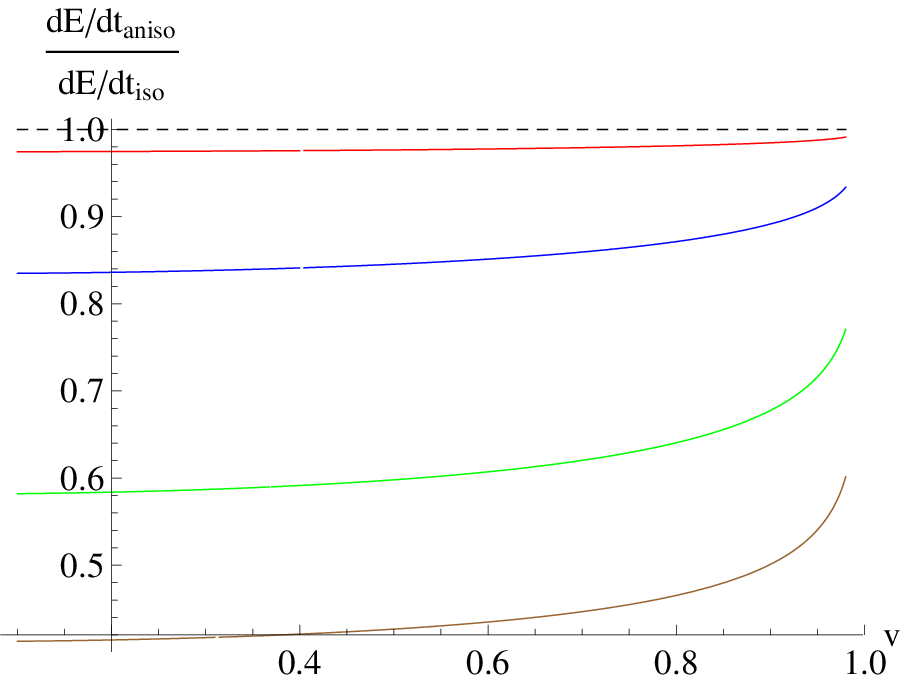}\includegraphics[width=2in]{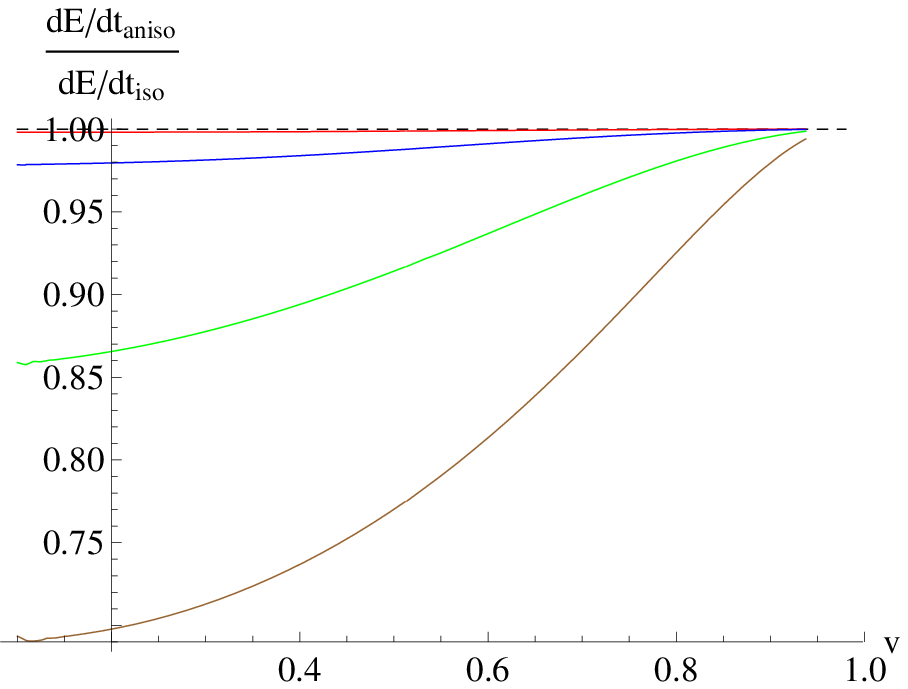}\includegraphics[width=2in]{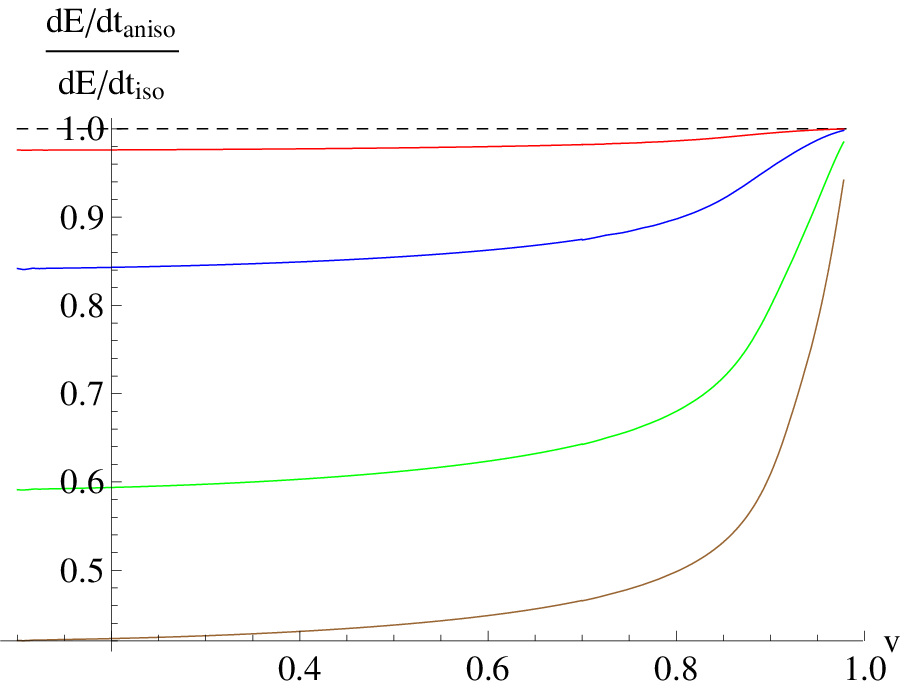}}
\caption{The ratio of the energy loss in anisotropic backgrounds to the isotropic cases for different angular velocities:  $\omega=0.05, 0.5, 5.0$ from left to right. }\label{e-loss-v}
\end{figure}

From the boundary theory point of view, to compare the energy loss
in two different media it is natural to fix the velocity,
$v=l\omega$. The ratios of the  energy loss rate in anisotropic
background to the isotropic case as a function of the velocity of
the rotating quark are given in Fig.\ref{e-loss-v}. These plots are
our main outputs and in following sections we will investigate these
results both analytically and numerically.

We summarize the important features of the plots in Fig.\ref{e-loss-v} as follows:
\begin{itemize}
\item{Again one can see that the
ultra-relativistic quark (or large $\Pi$ from gravity point of view) can not see the anisotropy in terms of its
energy loss rate, except for small anisotropy and small angular velocity and the same temperature case.
As expected, this is very similar to the drag force studied in \cite{Chernicoff:2012iq}.}
\item{The ratios of energy loss rates  are monotonic
functions, except for $\omega=0.5$ and large enough $a/T$ and the
same temperature. In this exception one can see a competition for
energy loss between the linear drag and the radiation channels. We
will discuss  this point later. }
\item{For fixed velocity $v$ and anisotropy the bigger angular
velocity $\omega$ leads to the smaller ratio of energy loss rate,
except for large velocity which breaks for medium $\omega$. }
\end{itemize}

One should note that the linear drag is a function of velocity of
the quark while, the radiation is a function of angular
acceleration, $\hat{\alpha}=v\,\omega$. Since $v\leq1$, small enough
$\omega$ leads to  small $\hat{\alpha}$ and small radiation for any
velocity. Therefore for small $\omega$ the dominant part of the
energy loss is due to the drag force. One interesting situation is
for intermediate $\omega$. In this case for small $v$ the drag force
is dominant while for large enough $v$, $\hat{\alpha}$ is large
enough and there is a competition for energy loss between drag and
radiation. As one can see in Fig.\ref{e-loss-v}, this is an
interesting part for the energy loss of a quark in this anisotropic
plasma model.

\begin{figure}
\subfigure[same temperature]
{\centerline{\includegraphics[width=55mm]{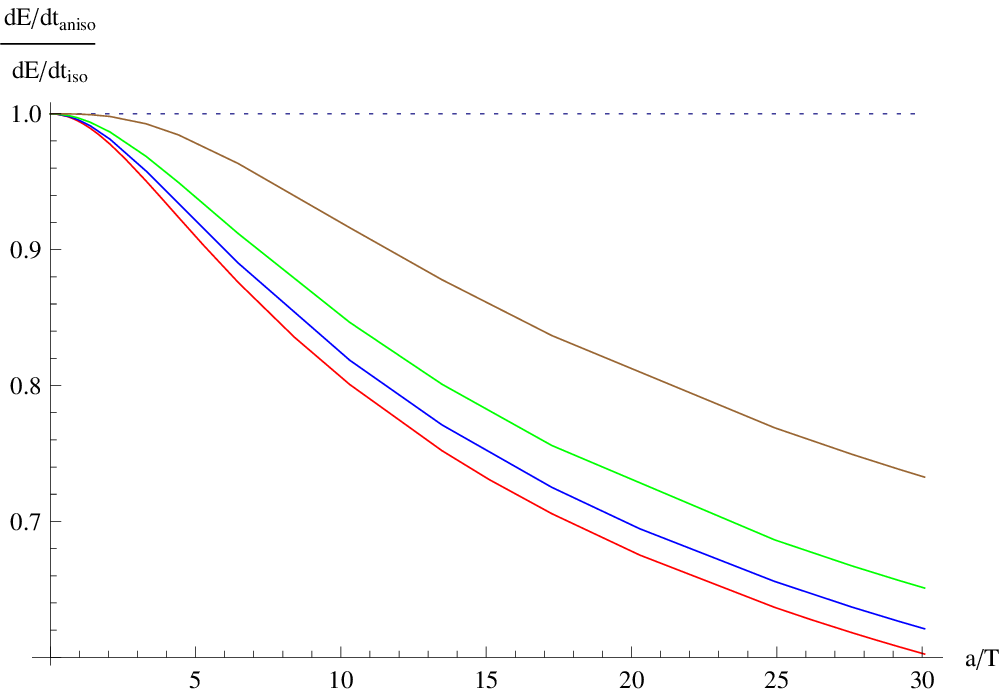}\includegraphics[width=55mm]{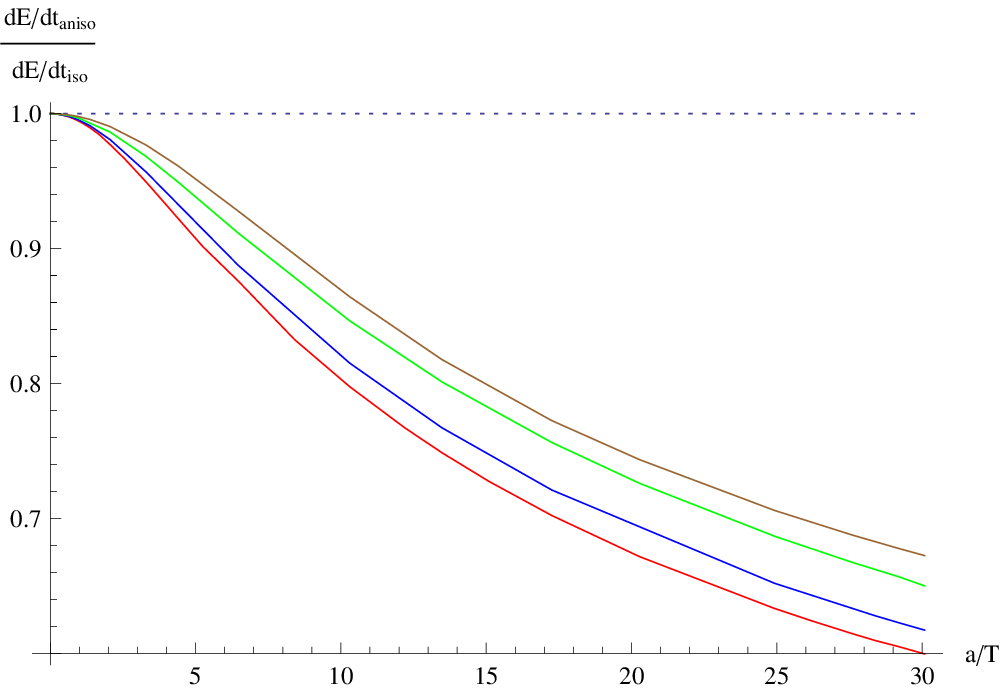}
\includegraphics[width=55mm]{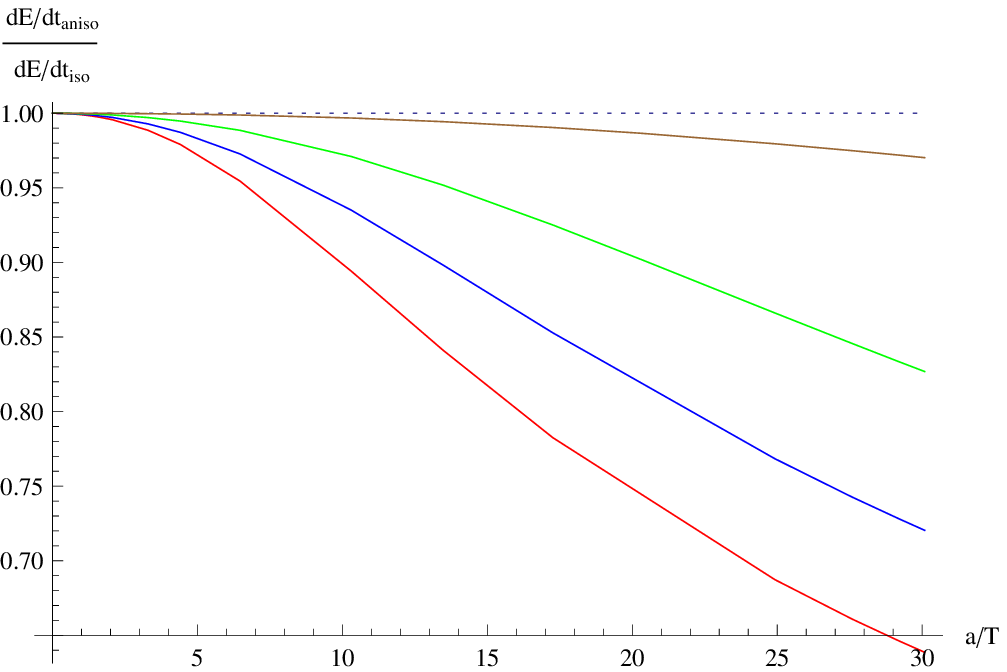}}}
\subfigure[same entropy and $A=N_c^{2/3}a/s^{1/3}$]
{\centerline{\includegraphics[width=55mm]{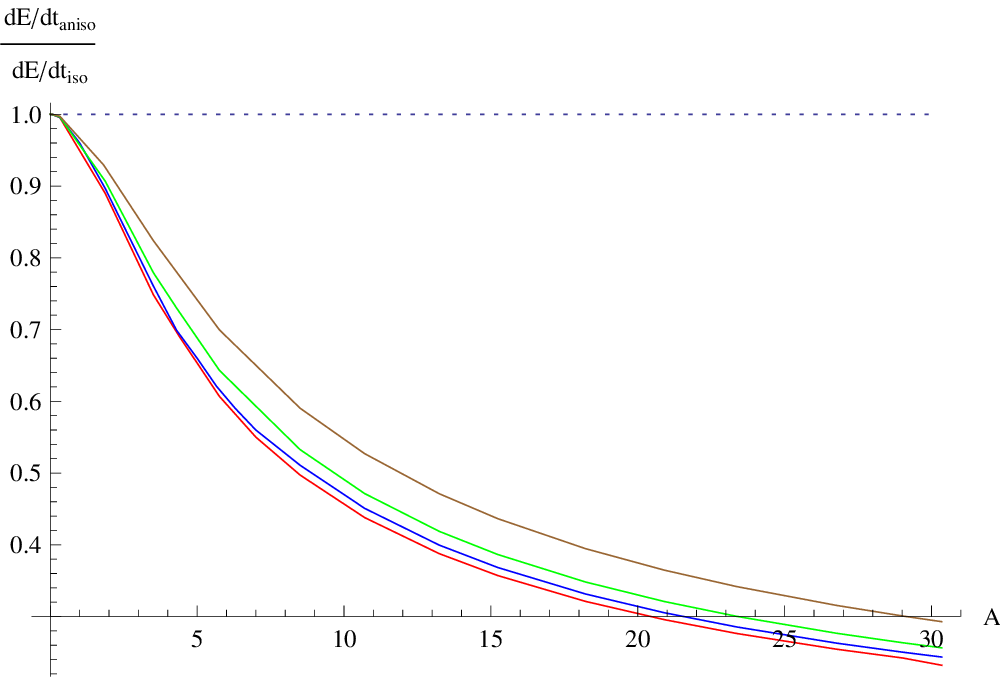}\includegraphics[width=55mm]{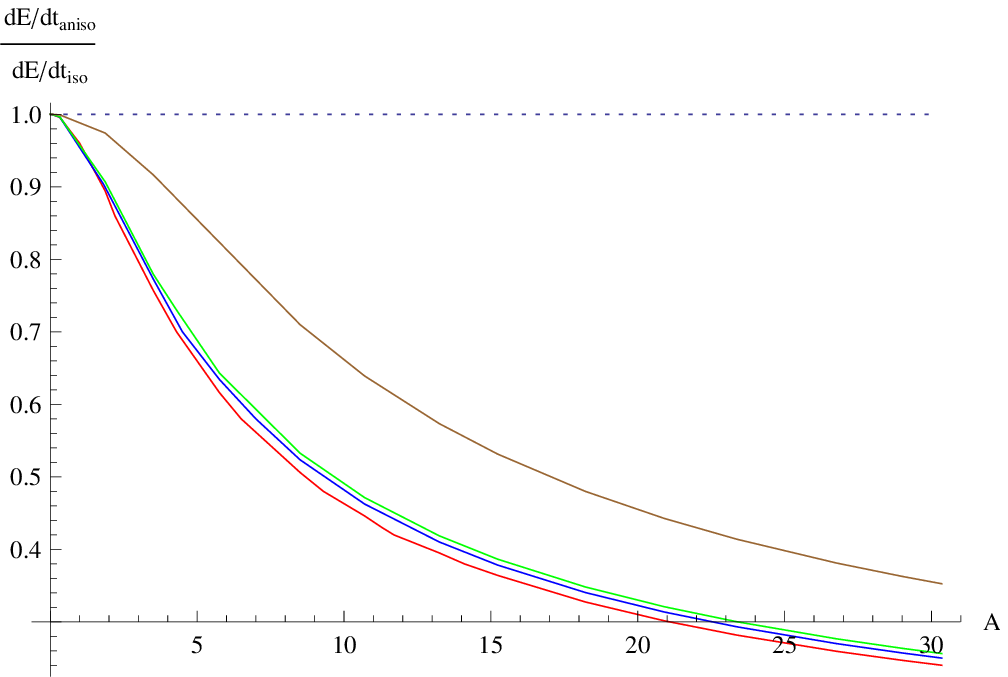}
\includegraphics[width=55mm]{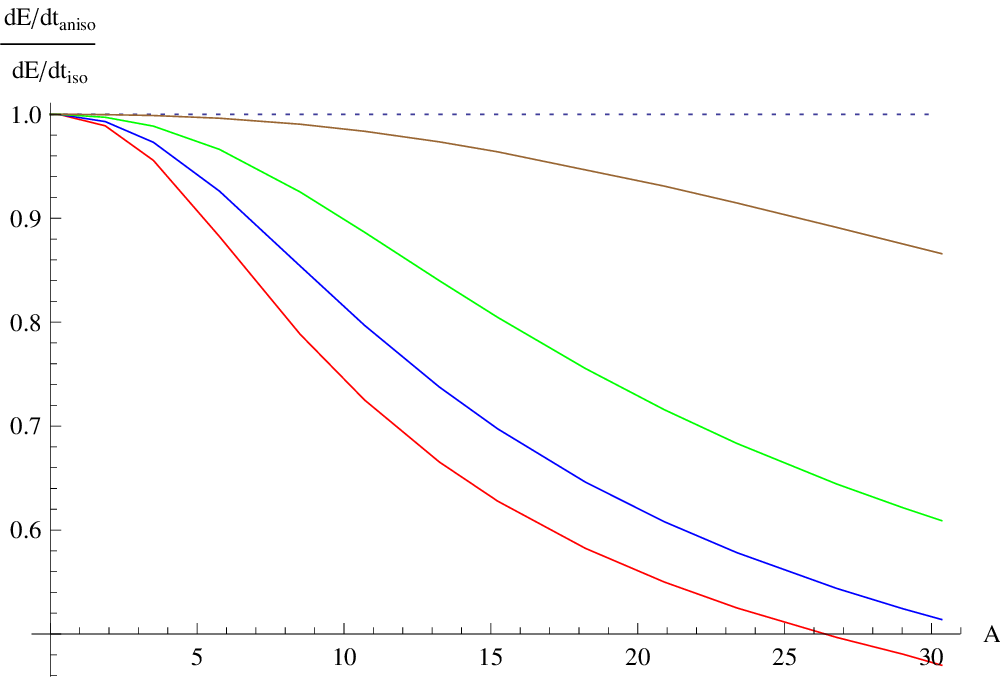}}}
\caption{Energy loss of rotating quark as a function of anisotropy
at the same temperature $(a)$ or at the same entropy density $(b)$
for four different velocities $v = 0.2, 05, 0.7, 0.9$,  from bottom to top in each
plot. In each arrow, angular velocities
are $\omega= 0.05, \omega= 0.5$ and $\omega= 5.0$ from left to
right.}\label{a-T-plaot}
\end{figure}

We studied the shape of rotating string in an anisotropic background
and the rate of energy loss of a rotating quark in the dual
anisotropic plasma, by using AdS/CFT correspondence, for some
specific values of anisotropy. In Fig. \ref{a-T-plaot}, we show the
ratio of energy loss rate of anisotropic to isotropic plasmas as
function of $a/T$ for four velocities. The first row is for the same
temperature cases and the second row is for the same entropy density
and in each row the angular velocity is $\omega=0.05, 0.5, 5.0$ from
left to right. From these plots it is obvious that the rate of
energy loss in anisotropic background is smaller than the isotropic
case and for fixed $v$ and $\omega$ the larger anisotropy leads to
the smaller rate of energy loss. This is the case both in the same
temperature and the same entropy cases.

It would be very interesting to ask about mechanism of energy loss
in two plasmas, is it drag force or radiation? In section \ref{cwldasr},
by comparing the results with both linear drag and synchrotron radiation,
we will show that the anisotropy decreases the radiation when the parton
propagates perpendicular to the direction of the anisotropy. This is
in perfect agreement with result of \cite{Roy:2010zg}.

%---------------------------------------------------------------------------------------------------------------------------------------------------
\section{Analytic limits}\label{al}
In this section, we consider two different limits in which we can do some analytical calculations, ultra-relativistic
limit and small $a/T$ limit. For these cases we know the metric
functions of dual gravity (\ref{aniso-background}) explicitly
\cite{Mateos:2011ix} and we can get some analytical results to
justify our numeric results at least in these limits. The results
also show that the energy loss rate depend on the temperature (or
horizon radius) and $a/T$, as expected from the discussion at the end of section \ref{tgb}.

%------------------------------------------------------------
\subsection{Ultra-relativistic limit}\label{url}
There are two different options for an ultra-relativistic limit,%
\begin{itemize}
\item{Large but fixed $v=l\,\omega=\text{const.}$, $\hat{\alpha}=v\,\omega\rightarrow0$, meaning $\omega\rightarrow0$}
\item{Large but fixed $v=l\,\omega=\text{const.}$, $\hat{\alpha}=v\,\omega\rightarrow\infty$, meaning
$\omega\rightarrow\infty$}
\end{itemize}

For the second case, as we see in Fig.\ref{rho-u-temp} and
Fig.\ref{rho-u-entropy},  the critical point $u_c$ can be
anywhere between the horizon $u = u_H$ and the boundary $u = 0$. For
this case which is completely general we know nothing analytically.
But for the first case the critical point $u_c$ approaches to the
boundary $u_c\rightarrow0$,\footnote{From Fig.\ref{rho-u-temp} and
Fig.\ref{rho-u-entropy}, it is easy to see that for fixed $\omega$,
a larger velocity correspond to a larger radius of rotation at the
boundary which leads to a larger $\Pi$ and a smaller $u_c$.}
which is similar to the drag force in the large velocity
limit $v\rightarrow1$ \cite{Chernicoff:2012iq}.
 Therefore one can use
the near boundary expansion of metric functions \cite{Mateos:2011ix},
\bea
&&\mathcal{F}=1+\frac{11a^2}{24}u^2+\left(\mathcal{F}_4+\frac{7a^4}{12}\log u\right)\,u^4+\mathcal{O}(u^6),\nn\\
&&\mathcal{H}=1+\frac{a^2}{4}u^2-\left(\frac{2\mathcal{B}_4}{7}-\frac{5a^4}{4032}-\frac{a^4}{6}\log u\right)\,u^4+\mathcal{O}(u^6),\label{uv-analitic}\\
&&\mathcal{B}=1-\frac{11a^2}{24}u^2+\left(\mathcal{B}_4-\frac{7a^4}{12}\log
u\right)\,u^4+\mathcal{O}(u^6),\nn
\eea
where
$\mathcal{F}_4$ and $\mathcal{B}_4$ depend on $a$, $T$ and a conformal anomaly scale, but they are
not determined by the near boundary analysis \cite{Mateos:2011ix}.

To find the critical point ($u_c, \rho_c$) for this case, we solve
the equations (\ref{critical-aniso}) analytically using the above
expansion of metric functions \eqref{uv-analitic}. The second
equation in (\ref{critical-aniso}) simplifies to \be
\Pi\,\omega=\frac{v_c^2}{u_c^2},\label{uc-vc} \ee where
$v_c\equiv\rho_c\,\omega$ is the velocity of the string at the
critical point, $u_c$. Note that in this limit the radius of the
rotating string on the boundary, $l$, is very close to the radius of
the string at the critical point, $\rho_c$, Fig.\ref{rho-u-temp} and
Fig.\ref{rho-u-entropy}. Therefore the velocity of the string at
these two points are also very close to each other. By using this
point, we can find the value of $u_c$ by solving the first equation
in (\ref{critical-aniso}) to leading order in $1-v_c^2$, \bea
\frac{1}{u_c^2}=\frac{T^2}{\sqrt{1-v_c^2}}\,\Xi,\hspace{20mm}\Xi=\sqrt{\frac{121a^4}{576T^4}-\frac{\mathcal{F}_4+\mathcal{B}_4}{T^4}}
\eea which is the same expansion for the critical point found for
the drag force calculation \cite{Chernicoff:2012iq}. But we should
remind that $v_c$ is a function of $u_c$ via eq.(\ref{uc-vc}) so the
explicit relation for $u_c$ is given by \be
u_c=\frac{1}{\Xi\,T^2}{\sqrt{-\frac{\Pi \omega
}{2}+\frac{1}{2}{\sqrt{\Pi^2 \omega ^2+4 \,\Xi^2 T^4}}}}, \ee

Note that the energy loss rate is given by the time component of the
background metric at the critical point, $|G_{tt}(u_c)|$
eq.(\ref{energy-loss}) and is given by \be
\frac{dE}{dt}\bigg|_{aniso}=\frac{\sqrt{\lambda}T^2}{2\pi}\frac{v_c^2}{\sqrt{1-v_c^2}}{\sqrt{\frac{121
a^4}{576T^4}- \frac{\mathcal{F}_4+\mathcal{B}_4}{T^4}}
}.\label{eloss-aniso} \ee Again, this is exactly the same results
which was found for the linear drag in \cite{Chernicoff:2012iq}.

For the linear drag, there is only one velocity for all of the points
of string but in our case the velocity is the velocity of the
critical point of the string, $v_c$. As it is shown in
Fig.\ref{rotate-drag} for small enough $v$, corresponds to
small enough $\Pi$, for
$\omega=0.05$ and $\omega=0.5$ the dominant part of the energy loss is due to the linear drag.
But the validity for this approximation is not
just $\hat{\alpha} \rightarrow 0$.
As one can see in Figs.\ref{rho-u-temp}  and \ref{rho-u-entropy}
for large $\Pi$  the radii of rotating string on the
boundary $u=0$ and the critical point $u_c$ are very close to each other  while,
the rest of the string below the critical point has larger radius.
Therefore, the $v\simeq v_c$ approximation is not valid any more for all of the string points.
This limit is discussed in \cite{Fadafan:2008bq} for the isotropic plasma as well.

Since the the energy loss rate (\ref{eloss-aniso}) is the same as
the linear drag results given in \cite{Chernicoff:2012iq} and its validity
range is as same as isotropic case, it's clear that the
comparison between the energy loss rate in the anisotropic and in the isotropic
backgrounds leads to the same results found for the drag force in \cite{Chernicoff:2012iq}.
Remind that this is true just in special limit,
large and fixed $v=l\,\omega$,  $\omega\rightarrow0$ and small enough $l$.

For small and large $a/T$ limits the $\mathcal{F}_4$ and $\mathcal{B}_4$ parameters are known \cite{Mateos:2011ix}
and one can find more explicit results.
For small $a/T$,
\bea
&&\mathcal{F}_4=-\pi^4T^4-\frac{9\pi^2T^2}{16}a^2-\left[\frac{101}{384}-\frac{7}{12}\log\left(\frac{2\pi T}{a}\right)-\frac{7}{12}\log\left(\frac{a}{\Lambda}\right)\right]\,a^4+\mathcal{O}(a^6),\nn\\
&&\mathcal{B}_4=\frac{7\pi^2T^2}{16}a^2+\left[\frac{593}{1152}-\frac{7}{12}\log\left(\frac{2\pi
T}{a}\right)-\frac{7}{12}\log\left(\frac{a}{\Lambda}\right)\right]\,a^4+\mathcal{O}(a^6),
\eea where $\Lambda$ relates to the conformal anomaly scale. It is
easy to see that the ratio of the energy loss in the anisotropic
plasma to the the energy loss in the isotropic case doesn't depend
on parameter $\Lambda$. For the same temperature, one can show that
this ratio is given by \be
\frac{dE/dt_{aniso}}{dE/dt_{iso}}\bigg|_{\text{temp}}=1+\frac{a^2}{16\pi^2T^2}
+\mathcal{O}\left(\frac{a^4}{T^4}\right).\label{eaniso-eiso-small}
\ee%
Therefore the energy loss rate of ultra-relativistic quark rotating
uniformly in transverse direction plane in an anisotropic plasma
with small $a/T$, small $\omega$ and large but fixed $v$ is greater
than the isotropic case at equal temperatures. This is similar to
the drag force results for small $a/T$ and large but fixed $v$
\cite{Chernicoff:2012iq}. This analytic result is in agreement with
our numerical results given in Fig.\ref{laniso-liso-temp} (red
curves in the first row plots) and Fig.\ref{e-loss-v} (red and blue
curves in the top-left plot) .

To compare the results for equal entropy density, we use the
relation between entropy density and temperature for an anisotropic
plasma in small $a/T$ \cite{Mateos:2011ix} \be
s=\frac{\pi^2N_c^2T^3}{2}+\frac{N_c^2T}{16}a^2+\mathcal{O}\left(\frac{a^4}{T^4}\right),
\ee which causes to different ratio for the rate of energy loss for
the same entropy densities, \be
\frac{dE/dt_{aniso}}{dE/dt_{iso}}\bigg|_{\text{entropy}}=1-\frac{N_c^{4/3}}{48(2\pi)^{3/2}}\left(\frac{a}{s^{1/3}}\right)^2+\mathcal{O}\left(\frac{a^4}{s^{4/3}}\right).\label{eaniso-eiso-small}
\ee Again, it is similar to the drag force results and the energy
loss rate in the anisotropic plasma is smaller than the isotropic
case at equal entropy density. These analytic results are in
agreement with our numerical results given in
Fig.\ref{laniso-liso-ent} (red curves in the first row plots) and
Fig.\ref{e-loss-v} (red and blue curves in the down-left plot) .

As expected, for large $a/T$ the results are the same as
the linear drag and the energy loss in the
anisotropic plasma is always greater than the energy loss in the isotropic case both
for equal temperatures and equal entropy densities. Using the
explicit relations for $\mathcal{F}_4$ and $\mathcal{B}_4$
parameters \cite{Mateos:2011ix}
\bea
\mathcal{F}_4=\frac{1}{132}\left[132 a^4 c_{\text{int}}+77 a^4 \log\left(\frac{a}{\Lambda}\right)-384c_{\text{ent}}\pi^2 a^{1/3} T^{11/3}+\cdots\right],\\
\mathcal{B}_4=\frac{1}{6336}\left[1331 a^4-6336a^4 c_{\text{int}}-3696 a^4 \log\left(\frac{a}{\Lambda}\right)+4032c_{\text{ent}}\pi^2 a^{1/3} T^{11/3}+\cdots\right],
\eea
the ratio of the energy loss rate in the anisotropic plasma to the energy loss rate in the isotropic case is given by
\bea
&&\frac{dE/dt_{aniso}}{dE/dt_{iso}}\bigg|_{\text{temp}}=\frac{\sqrt{2c_{\text{ent}}}}{\pi}\left(\frac{a}{T}\right)^{1/3}+\cdots,\\
&&\frac{dE/dt_{aniso}}{dE/dt_{iso}}\bigg|_{\text{entropy}}=\frac{1}{2^{1/6}\pi^{1/3}c_{\text{ent}}^{3/16}N_c^{1/24}}\left(\frac{s^{1/3}}{a}\right)^{1/16}+\cdots,
\eea for equal temperatures and equal entropy densities,
respectively. $c_{\text{ent}}$ is a constant introduced in the
entropy density in section \ref{tgb} and  $c_{\text{int}}$ is an
integration constant introduced in \cite{Mateos:2011ix}. As
expected, this is again the same results as drag force
\cite{Chernicoff:2012iq}.

%-----------------------------------------------------------
 \subsection{Small $a/T$ limit}\label{sal}
For small $a/T$ the metric functions and the radius of the horizon are known as some expansions around the black D3-brane solution  \cite{Mateos:2011ix},
\bea
&&\mathcal{F}=1-\frac{u^4}{u_H^4}+a^2\mathcal{F}_2+\mathcal{O}(a^4),\nn\\
&&\mathcal{B}=1+a^2\mathcal{B}_2+\mathcal{O}(a^4),\nn\\
&&\log\mathcal{H}=\frac{a^2u_H^2}{4}\log\left(1+\frac{u^2}{u_H^2}\right)+\mathcal{O}(a^4),\nn\\
&&u_H=\frac{1}{\pi T}+\frac{5\log 2-2}{48\pi^3T^3}a^2+\mathcal{O}(a^4),\label{small-a-T}
\eea
in which
\bea
&&\mathcal{F}_2=\frac{1}{24u_H^2}\left[8u^2(u_H^2-u^2)-10u^4\log2+3u_H^4+7\log\left(1+\frac{u^2}{u_H^2}\right)\right],\\
&&\mathcal{B}_2=-\frac{u_H^2}{24}\left[\frac{10u^2}{u_H^2+u^2}+\log\left(1+\frac{u^2}{u_H^2}\right)\right].
\eea It is easy to find the critical point by using the general
formula (\ref{critical}) and above expansions for the background
metric. Obviously, the result is an expansion around the isotropic
critical point (\ref{critical-iso}). Up to second order in $a$ the
critical point is given by \be
u_c=u_{\text{iso}}+a^2u_{c}^{(2)}+\mathcal{O}(a^4),\hspace{20mm}\rho_c=u_c\sqrt{\frac{\Pi}{\omega}},\label{uc-small-a-T}
\ee where \bea
u_{\text{iso}}=\frac{1}{\pi^2T^2}\sqrt{-\frac{\Pi\omega}{2}+\frac{1}{2}\sqrt{\Pi^2\omega^2+4\pi^4T^4}},\nn\\
\eea Doing some calculations it is easy to show that the first
correction to the isotropic critical point is \bea
u_{c}^{(2)}=\frac{3-2 \pi ^2 T^2 u_{\text{iso}}^2(1+\pi ^2
T^2u_{\text{iso}}^2)+\left(-1+8 \pi ^4 T^4u_{\text{iso}}^4\right)
\log\left(1+\pi ^2 T^2u_{\text{iso}}^2\right)}{48 \pi ^2
T^2u_{\text{iso}} \left(2 \pi ^4 T^4 u_{\text{iso}}^2+ \Pi  \omega
\right)}. \eea%
In contrast to the ultra-relativistic limit studied in the previous
subsection, one can not find the analytic results from boundary they
point of view. It's because, there is no analytic relation between
the radius of rotation at the critical point $\rho_c$ and the radius
of rotation at the boundary $l$, which is the case for
ultra-relativistic limit (see the discussion after
eq.(\ref{uc-vc})). Therefore to explain every thing from boundary
theory point of view we must solve the general equation of motion
for $\rho$ (\ref{eom-rho}) numerically from critical point to the
boundary by using the explicit metric functions \eqref{small-a-T}
and the explicit relation for the critical point
\eqref{uc-small-a-T}. The results are in perfect agreement with the
results we found numerically for small $a/T$ in previous section and
since there is no more point we don't show the plots here.
%---------------------------------------------------------------------------------------------------------------------------------------------------
\section{Comparison with the linear drag and the synchrotron radiation}\label{cwldasr}

In this section we compare our results of the rate of energy loss
for a rotating quark in an anisotropic plasma with two different
known cases, the linear drag and the synchrotron radiation. First we
want to compare the energy loss rate of rotating quark with the
energy loss rate in the linear drag case. The drag force experienced
by an infinitely massive quark propagating at constant velocity
through an anisotropic, strongly-coupled $\mathcal{N} = 4$ SYM
plasma was studied in \cite{Chernicoff:2012iq, Giataganas:2012zy}.
\begin{figure}
\subfigure[$\omega=0.05$]{\includegraphics[width=52mm]{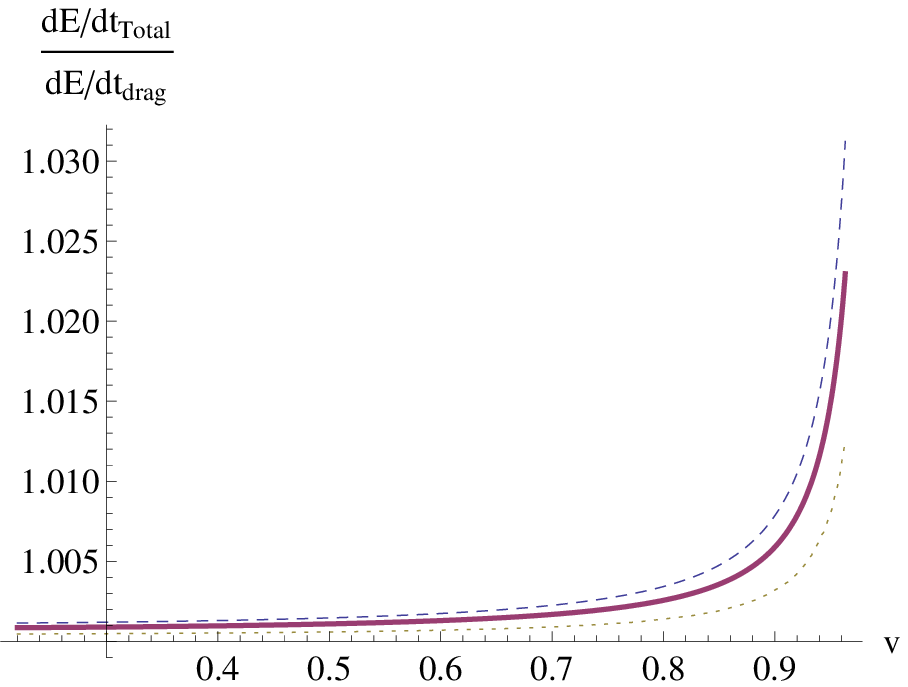}}
\subfigure[$\omega=0.5$]{\includegraphics[width=50mm]{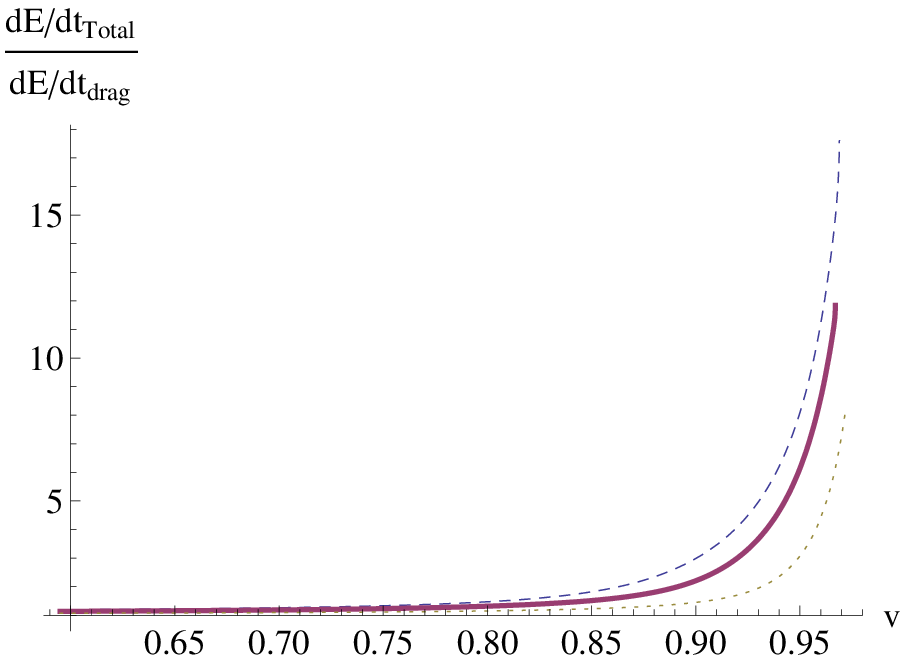}}
\subfigure[$\omega=5.0$]{\includegraphics[width=52mm]{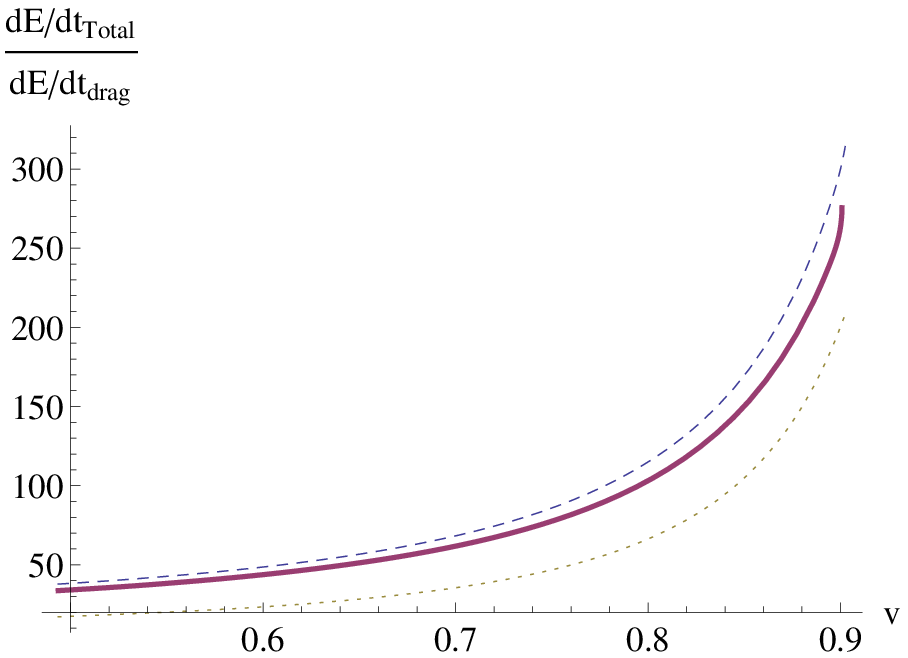}}
\caption{The ratio of the total energy loss rate of
rotating quark to its linear drag part as a function of velocity for  $a/T\simeq$1.4 (dashed), 12(solid), 86 (dotted) curves.}
\label{rotate-drag}
\end{figure}
In
Fig.\ref{rotate-drag} we show the ratio of the total energy loss rate of
rotating quark to its linear drag part as a function of velocity of the
quark in the boundary theory for different $\omega$ and different $a/T$.

As expected, for small $\omega$ the dominant part of the ratio of
energy loss is due to the linear drag for a wide range of velocity of the
quark on the boundary. In this range, the local
velocity of the rotating string at each point can be approximated by the velocity of the
quark at the boundary which means the string behaves
similar to the linear drag case studied in \cite{Chernicoff:2012iq, Giataganas:2012zy}.

In this limit we show a perfect agreement between the energy loss rate of rotating quark and the linear drag analytically.
The rate of energy loss for a rotating string is given by (\ref{energy-loss}) or equivalently by
\be
\frac{dE}{dt}=\frac{\Pi\omega}{2\pi\alpha'}=\frac{\Pi v_c}{2\pi\alpha'
l}.
\ee
Note that in this limit the velocity of string is close to $v_c$ for all of the points.
Using well-known relations between rotation and linear
motion, constant velocity at each point of the rotating string and
equation (\ref{el-drag}) one can see that the rate of energy loss is given by,
\be
\frac{dE}{dt}=\frac{\Pi\omega}{2\pi\alpha'}=\frac{\mathcal{P} v}{2\pi\alpha'}=\frac{dE^{(\text{drag})}}{dt}.
\ee
This result is broken if the radius of the rotation for
$u_H>u\geq u_c$ ($\rho(u)>\rho_c$) is larger than $\rho_c$, which means the constant velocity for
all of the string's points is not good approximation.
This is the case for small $\omega$ but very large $\Pi$, discussed also in earlier sections.

It is easy to see that for larger $\omega$  the dominant part of the rate of energy loss is due to
the drag only for small velocities  of the quark.
As expected, in this limit the dominant part of the energy loss is due to the  radiation for large enough velocities of the quark.

In \cite{Fadafan:2008bq}, it was shown that for large $\omega$ and
large enough velocities of the quark, the dominant part of the
energy loss is due to the radiation channel for isotropic
$\mathcal{N}=4$ SYM plasma. This result was shown by comparing the
energy loss with the synchrotron radiation introduced by Mikhailov
\cite{Mikhailov:2003er}. So it is natural to compare the rate of
energy loss with the synchrotron radiation using the Mikhailov's
formula, which is the radiation of a quark rotating in a vacuum, \be
\frac{dE}{dt}|_{\text{v.
r.}}=\frac{\sqrt{\lambda}}{2\pi}\frac{v^2\omega^2}{(1-v^2)^2}.\label{mikh-formula}
\ee

\begin{figure}
\subfigure[$\omega=0.05$]{\includegraphics[width=52mm]{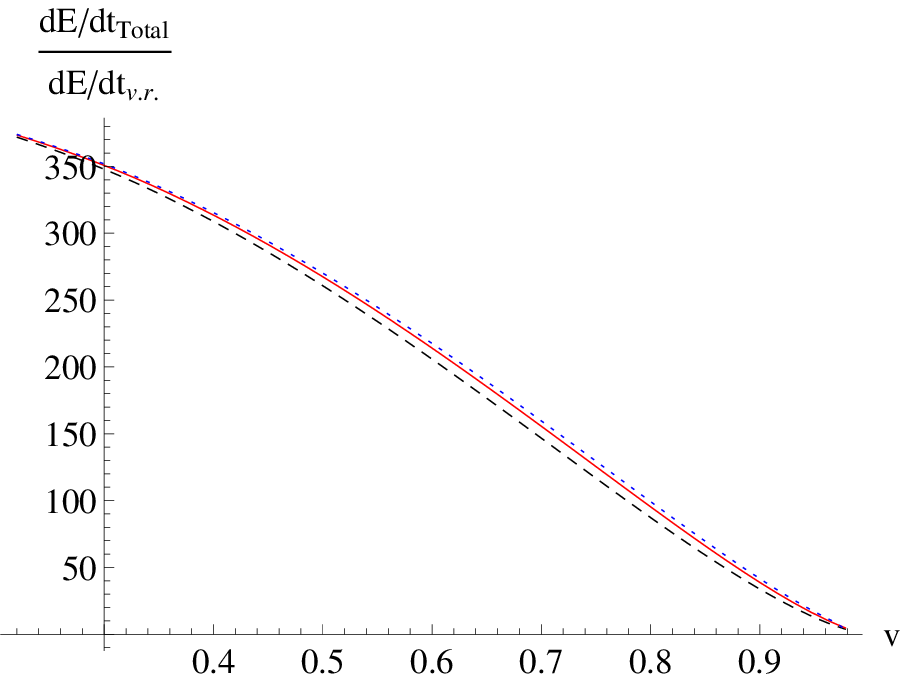}}
\subfigure[$\omega=0.5$]{\includegraphics[width=50mm]{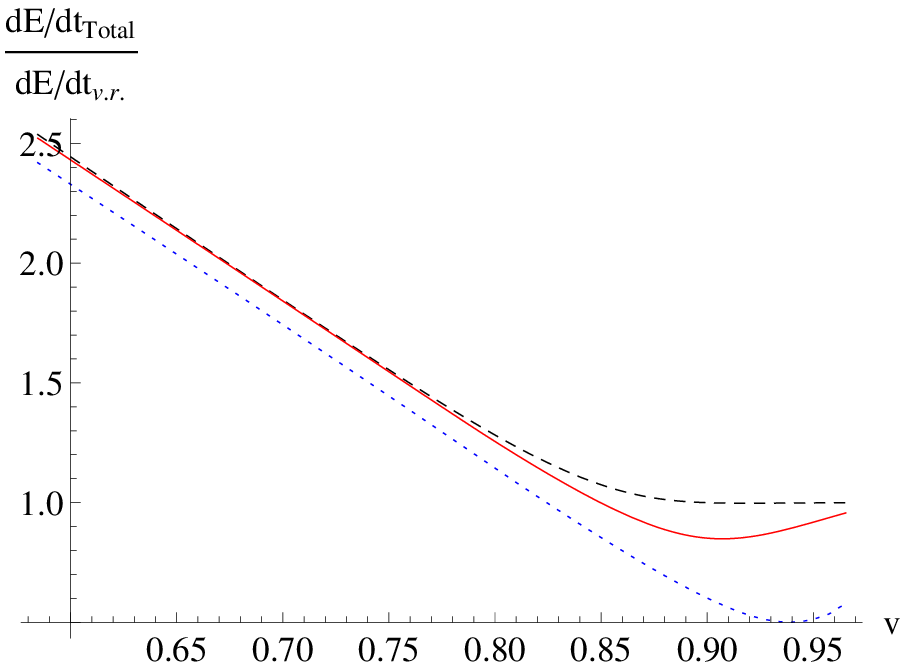}}
\subfigure[$\omega=5$]{\includegraphics[width=52mm]{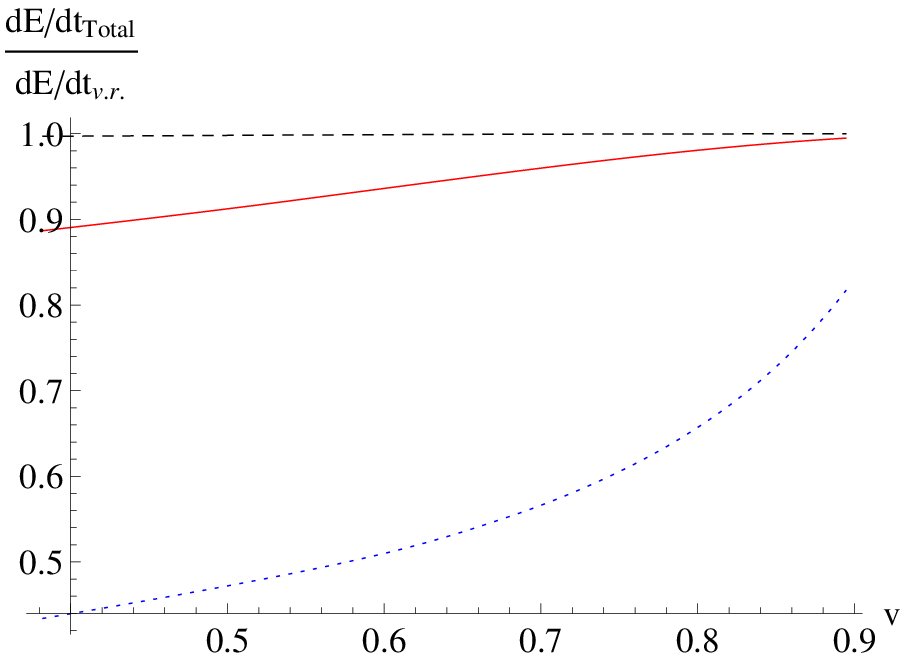}}
\caption{The ratio of the rate of energy loss to the radiation.
$a/T\simeq$0.8(dashed), 4.4(solid), 86(dotted) curves.}\label{rotate-radiate}
\end{figure}

Since this formula can describe the radiation in isotropic plasma
\cite{Fadafan:2008bq}, the ratio of the rate of energy loss
\eqref{energy-loss} to the above rate \eqref{mikh-formula}  is
actually the ratio of total energy loss to the radiation in
isotropic plasma. The results are given in Fig.
\ref{rotate-radiate}. For small $\omega$ one can see that the total
energy loss rate is much more than the radiation part, while in Fig.
\ref{rotate-drag} we show that the dominant part of the energy loss,
in this case, is due to the linear drag.

For large $\omega$ and large enough velocity of the quark in
Fig.\ref{rotate-radiate}, where the dominant part of energy loss is
due to the radiation, one can see that the radiation in anisotropic
plasma is less than isotropic case and a larger $a/T$, causes to a
smaller radiation. For fixed $a/T$, a larger velocity of the quark
leads to a smaller ratio of the energy loss rate to the vacuum
radiation such that for ultra-relativistic quarks with large
$\omega$, the energy loss is closer to the vacuum radiation.

Again, one interesting range is intermediate angular velocity, $\omega=0.5$, and large $v$.
In this regime one can see the effect of anisotropy on the competition between the drag and radiation channels in energy loss.
This behavior is similar to the first row of Fig.\ref{e-loss-v}. Naively one can conclude that comparing the energy loss in the anisotropic and isotropic plasma at equal temperatures is more natural than at equal entropy densities.

%---------------------------------------------------------------------------------------------------------------------------------------------------
\section{Summary and discussion}\label{sad}
%-----------------------------------------------------------------------------------------------------
In this paper, we have studied the energy loss of a rotating heavy
quark in a strongly-coupled anisotropic plasma from holography.
There are two models to investigate the anisotropic plasma from
AdS/CFT correspondence which have been proposed by Mateos and
Trancanelli in \cite{Mateos:2011ix} and by  Janik and Witaszczyk in
\cite{Janik:2008tc}.\footnote{Other AdS backgrounds which are dual
to an anisotropic fluids are constructed in \cite{Erdmenger:2011tj}}
In JW model \cite{Janik:2008tc}, a geometry involving a
comparatively benign naked singularity was considered, while MT
model \cite{Mateos:2011ix} has regular geometry involving a
nontrivial axion field dual to a parity-odd deformation of the gauge
theory. Different aspects of these models have been studied in
\cite{Rebhan:2012bw}. The latter model has been used in this paper.
In this model, the anisotropic $\mathcal{N}=4$ SYM plasma is
symmetric in $xy$ plane but in the beam direction ($z$ direction) is
not. Then in the context of heavy ion collisions, we considered a
rotating test quark within the transverse plane which is also mostly
interested in jets.

Holographic study shows different momentum broadening in an anisotropic
medium along the beam axis and along the transverse plane
\cite{Chernicoff:2012iq,Giataganas:2012zy,Chernicoff:2012gu}.
These quantities in a weakly coupled anisotropic plasma have been studied
in \cite{Romatschke:2004au,Romatschke:2006bb,Dumitru:2007rp}.
It was shown that, the momentum broadening along the beam axis increases
slightly in the presence of anisotropy, while the momentum
broadening in the transverse plane decreases more significantly.
The jet quenching parameter was studied in
\cite{Giataganas:2012zy,Chernicoff:2012gu,Rebhan:2012bw} and it was
found that for large anisotropic parameter the same result can be
achieved. We showed that in the regime where we expect to see the
radiation of beam of synchrotron-like radiation, the energy loss is
significantly less than the isotropic case.

We find valuable results by comparing of the anisotropic case with
the isotropic background at the same temperature or at the same
entropy density. The ratio of the energy loss rate in anisotropic
background to the isotropic case as a function of the velocity of
the rotating quark is given in Fig. 7. One finds that the energy
loss of ultra-relativistic heavy quark is the same as the drag force
calculations in \cite{Chernicoff:2012iq}. Also in Fig. 8, this ratio
as a function of a/T for different velocities of rotating quark is
shown. It is clearly seen that the rate of energy loss in
anisotropic background is smaller than the isotropic case and for
fixed $v$ and $\omega$ the larger anisotropy leads to the smaller
rate of energy loss.

It would be interesting to compare our results with the radiation
energy loss of heavy quarks in a real anisotropic plasma. In a
first-order opacity expansion it is studied in \cite{Roy:2010zg}. It
was found that the energy loss due to the radiation depends on the
direction of propagation of the fast partons with respect to the
anisotropy axis as well as on the anisotropy parameter. It was shown
that along parallel the anisotropy direction the radiation increases
while in the transverse direction it decreases. Regarding our setup,
we have considered the radiation in the transverse plane and we
found that the radiation in a strongly-coupled anisotropic plasma is
less than the isotropic case.

The energy density and angular distribution of the power radiated by
a rotating quark in a strongly-coupled isotropic plasma was studied
in \cite{Athanasiou:2010pv}. This study was also extended to the
case of non-zero temperature in \cite{Chesler:2011nc}. It would be
interesting to study effect of anisotropy on the angular
distribution of the power radiated by a heavy quark rotating in an
anisotropic plasma. We showed that the crossover from the
drag-dominated regime to the radiation-dominated regime
significantly depends on the anisotropic parameter. It was found
that unlike in a vacuum, in the plasma at nonzero temperature the
energy disturbance created by the rotating quark can excite in two
qualitatively distinct modes in the energy density; a sound mode and
a light-like mode which propagates at the speed of light. The energy
loss due to unstable modes in a weakly coupled anisotropic plasma
has been studied in \cite{Carrington:2011uj,Kurkela:2011ti}. It
would be very interesting to understand the effect of instability
modes on the energy density in an anisotropic strongly-coupled
plasma. This investigation can shed new light to the similarities
between the quenching of the beam of strongly-coupled synchrotron
radiation in the strongly-coupled anisotropic $\mathcal{N} = 4$ SYM
plasma and the quenching of jets in heavy ion collisions at the LHC
and RHIC.\\

%---------------------------------------------------------------------------------------------------------------------------------------------------
{\large  Acknowledgements}\\
 We would like to thank M.~Ali-Akbari, N.~Abbasi,  A.~Davody, M.~Abedini, E.~Azimfard, D.~Giataganas, K.~Goldstein D.~Nickel and M.~M.~Sheikh-Jabbari for useful discussions on the different aspects of rotating string in an anisotropic background.
 We are also very grateful to thank to K.~Goldstein and M.~Sohani for carefully reading the draft.
 K.~B.~F. thanks a lot from M.~Strickland for discussion on the papers in the subject of weakly anisotropic plasma.
 H.~S. thanks R. Warmbier and R. Morad for their comments on numerical calculations.
 We would like to thank the referee of JHEP for giving constructive comments which helped improving the paper.
 H.~S. would like to thank Institute for Research in Fundamental Sciences (IPM) for hospitality during different stages of preparing this project.
 H.~S. is supported by National Research Foundation (NRF). Any opinion, findings and
 conclusions or recommendations expressed in this material are those
 of the H.~S. and therefore the NRF do not accept any liability with
 regard thereto.

%---------------------------------------------------------------------------------------------------------------------------------------------------
%---------------------------------------------------------------------------------------------------------------------------------------------------
\appendix
%---------------------------------------------------------------------------------------------------------------------------------------------------
%---------------------------------------------------------------------------------------------------------------------------------------------------
 \section{Rotation in anisotropic plane}\label{riap}
 In this appendix we study a rotating quark in a general three dimensional spatial of an  anisotropic background (\ref{aniso-background}).
 Without loss of generality, we assume that the rotation is in $xz$-plane (extension to more general case is straightforward).
 Using the following polar coordinate transformation
 \be
 x=\rho \cos\psi,\hspace{20mm}z=\frac{\rho}{\sqrt{\mathcal{H}}}\sin\psi,
 \ee
 which is useful to study a rotating object, one can show that the $xz$ part of the background metric \eqref{aniso-background} is given by
 \be
 dx^2+\mathcal{H}~dz^2=d\rho^2+\rho^2 d\psi^2+\frac{\rho^2\sin^2\psi}{4}\frac{{\mathcal{H}'}^2}{\mathcal{H}^2}~du^2
 +{\rho\sin^2\psi}\frac{{\mathcal{H}'}}{\mathcal{H}}~du d\rho-\frac{\rho^2 \sin2\psi}{2} \frac{\mathcal{H}'}{\mathcal{H}}~du d\psi.
 \ee
 It's obvious  that there is no SO(2) symmetry in $xz$-plane for $\mathcal{H}\neq\text{constant}$. As we will see this has important consequence to find the energy loss rate.

 Using the usual ansatz for a rotating string in $xz$-plane,
 \be
 X^\mu(\tau, \sigma)=\left(t=\tau, u=\sigma, \psi=\omega t+\theta(\sigma), \rho=\rho(u), y=0\right),
 \ee
 it is easy to show that the Lagrangian density of Nambu-Goto action for a rotating string in $xz$-plane is given by
 \bea
 \mathcal{L}=\left[-(G_{uu}+G_{\rho\rho}{\rho'}^2+2G_{u\rho}\rho')(G_{tt}+G_{\psi\psi}\omega^2)
 -G_{tt}G_{\psi\psi}{\theta'}^2-G_{u \psi}(2G_{tt}\theta'-G_{u\psi}\omega^2)\right]^{1/2}.
 \eea
 Note that  the $G_{uu}, G_{u \rho}$ and $G_{u \psi}$ components of the background are functions of both $u$ and $\psi$ coordinates.
 Again, it shows that there is no SO(2) symmetry in this case.
In contrast to the rotation in $xy$-plane, studied in the main body of the paper,
not only $\theta'$ (the derivative of $\theta$ with respect to $u$), but also  $\theta$  appears explicitly in the action (via $\psi$ dependency),
\be
\mathcal{L}=\mathcal{L}(\rho, \rho', \theta, \theta').
\ee
 It means  the  momentum conjugate to the $\theta$ field is not a constant of motion anymore and a hanging string rotates in-homogeneously.

 On the other hand, from drag force investigations \cite{Chernicoff:2012iq} we know that for a string moving in $xz$-plane along nighter $x$ nor along $z$ directions, there is a misalignment between the velocity in the far infrared and the drag force $\overrightarrow{F}$. This is the main reason for difficulty of usual (holographic) calculation of a rotating quark in an anisotropic plane which leads to the inhomogeneous rotation explained above. In the other words, for a rotating string, while the end point of the string is moving on a circle at the boundary ($\mathcal{H}_b=1$), each point of the rest of the string is moving on an ellipse.
 It's interesting to solve the equations of motion of $\rho$ and $\theta$ fields (two coupled  second order differential equations) in anisotropic background for a general rotation in anisotropic background
 which is not in the scope of this paper.

%---------------------------------------------------------------------------------------------------------------------------------------------------
 \section{Linear drag in a general background}\label{ldigb}

In this appendix we give a brief review on a linear drag case corresponded to a quark moving on a line with a fixed velocity $v$ in a general background which has been known for some especial examples \cite{Gubser:2006bz}.
Using the static gauge for a string moving in $x$ direction of a general background (\ref{general-bg})
\be
X_{\text{drag}}^{\mu}(\tau,\sigma)=\left(t=\tau,\,u=\sigma,\,x=\xi(u)+v\,t,\,y=z=0\right),\label{ansatz-drag}
\ee%
one can show that the  Nambu-Goto density of lagrangian is given by
\be
\mathcal{L}_{\text{drag}}=\left[-G_{tt}G_{uu}-G_{xx}G_{tt}\xi'^2-G_{xx}G_{uu}v^2\right]^{1/2}.
\ee
Since $\xi$ does not appear explicitly in the action, its momentum conjugate is conserved and the equation of motion is given by
\be
\mathcal{P}=\frac{\rd \mathcal{L}}{\rd \xi'}=-\frac{G_{tt}G_{xx}\xi'}{\mathcal{L}}=\text{constant}.
\ee
Solving this equation for $\xi'$ yields to
\be
\xi'^2=-\mathcal{P}^2\frac{G_{uu}(G_{tt}+G_{xx}v^2)}{G_{xx}G_{tt}(G_{xx}G_{tt}+\mathcal{P}^2)}.
\ee
The positivity of LHS leads to the fact that there is a critical point, $u_c^{(\text{drag})}$, where both numerator and denominator change their signs and it is defined by
\bea
&&G_{tt}(u_c)^{(\text{drag})}+G_{xx}(u_c)^{(\text{drag})}v^2=0,\label{critical-drag1}\\
&&G_{xx}(u_c)^{(\text{drag})}\,G_{tt}(u_c)^{(\text{drag})}+\mathcal{P}^2=0,\label{critical-drag2}
\eea
which lead to
\bea
|G_{tt}(u_c)^{(\text{drag})}|=\mathcal{P}\,v,\hspace{20mm}G_{xx}(u_c)^{(\text{drag})}=\frac{\mathcal{P}}{v}.\label{critical-drag}
\eea
On the other hand, by using (\ref{critical-drag1}) and (\ref{critical-drag2}), it is easy to show that the rate of energy
loss due to the linear drag is given by
\be
\frac{dE^{(\text{drag})}}{dt}=\Pi^{\sigma(\text{drag})}_t=\frac{\mathcal{P}v}{2\pi\alpha'}=\frac{|G_{tt}(u_c)^{(\text{drag})}|}{2\pi\alpha'},\label{el-drag}
\ee
 where in the second equality we used the first equation in
(\ref{critical-drag}). Comparing equations (\ref{energy-loss}) and
(\ref{el-drag}) shows that the general form of the rate of energy
loss in both linear drag and rotating strings are the same and it's just
a function of the critical point.

Therefore, if the critical point at two different cases are
the same the rate of energy loss should be the same.
As we show in sections \ref{rsiscap} and \ref{al},
for small angular velocity of a rotating string the critical point is
very close to the linear drag both numerically and analytically.

% -------------------------------------------------------------------------------------------------------------------------
 % -------------------------------------------------------------------------------------------------------------------------

\end{document}